\documentclass[twocolumn]{aastex631}

\shorttitle{White Dwarf Masses For 208 Novae}
\shortauthors{Schaefer}
\graphicspath{{./}{figures/}}

\begin{document}
\title{Comprehensive Listing of 208 Nova White Dwarf Masses As the Primary Determinant of Spectral-Class and Light-Curve-Class}

\author[0000-0002-2659-8763]{Bradley E. Schaefer}
\affiliation{Department of Physics and Astronomy,
Louisiana State University,
Baton Rouge, LA 70803, USA}

\begin{abstract}

For Galactic novae, I calculate and collect a comprehensive catalog of 208 measures of white dwarf (WD) masses ($M_{\rm WD}$) and 232 measures of average $V$ magnitudes in quiescence ($V_q$).  These are collected into a comprehensive catalog of most fundamental properties of all 402 known Galactic novae.  The nova light curve and spectral classes are determined primarily by $M_{\rm WD}$.  With an apparently clean cutoff, nova with light curve shapes in the S, P, O, and C classes have $>$0.95 $M_{\odot}$, while the J, D, and F class novae have $<$0.95 $M_{\odot}$.  The speed class of the light curves is $t_3$=$10^{(-1.73M_{\rm WD})}$$\times$1900 days.  The spectral class of novae is Fe II below 1.15 $M_{\odot}$, is He/N above 1.15 $M_{\odot}$, and the Hybrid novae are spread around this division.  Neon novae have WD masses ranging from 0.53--1.37 $M_{\odot}$, with 76\% being measured to be below their minimum formation mass of 1.2 $M_{\odot}$, demonstrating that most are losing mass over each eruption cycle.  The FWHM velocity of the Balmer line profiles is close to 0.23 times the WD escape velocity, or roughly $10^{(M_{\rm WD}/2)}$$\times$500 km s$^{-1}$ for $<$1.3 $M_{\odot}$.  And all the known Galactic recurrent novae are $>$1.2 $M_{\odot}$.  For issues involving the late expansion of the ejecta, I find that the visibility of shells is strongly biased towards novae with orbital periods $<$0.33 days, and that the visibility of $\gamma$-rays from the shells are strongly biased towards novae with fast declines, with $t_3$ a proxy for the $\gamma$-ray luminosity.

\end{abstract}

\section{INTRODUCTION}

Novae are runaway thermonuclear eruptions on the surface of a white dwarf (WD) star in a close interacting binary involving a relatively normal star filling its Roche lobe and spilling gas onto the WD as mediated through an accretion disk (Payne-Gaposchkin 1964, Chomiuk, Metzger, \& Shen 2020).  Between eruptions, as the gas is accumulating on the WD surface, the quiescent nova appears as a cataclysmic variable (CV).  Most nova systems accumulate the mass needed to trigger the runaway eruption with recurrence time scales of 1--100 thousand years, while there are 11 known nova systems in our Galaxy that have recurrence times of under one century, with these being called recurrent novae (RNe).   Each eruption typically has a fast rise from quiescence of 1--10 days, a peak in the light curve lasting 1-100 days, and a slow fade back to minimum lasting 1--10 years or so.  As of mid-2022, 402 novae have been discovered in our Milky Way, while many more have exploded in the last century, only to be missed by nova hunters.

For the study of novae in quiescence, the most important property is the orbital period ($P$).  For the study of nova eruptions, the most important property is the WD mass ($M_{\rm WD}$).  Detailed studies of many individual novae is required, but the critical big-picture questions are all answered with demographics.  Historically, the big-picture questions addressed with nova demographics have included the existence of galaxies as `Island Universes' and measuring the Hubble Constant.  In recent years, a variety of questions (e.g., the hibernation of CVs, bulge versus disk populations, aluminum-26 production, the maximum magnitude versus rate of decline relation, and magnetic braking) have been largely solved from the properties of many novae considered together.  The forefront of the CV field is now dominated by questions with critical input from nova demographics.  The  all-important question of CV evolution comes down to discovering the source of the angular momentum loss (not magnetic braking), for which the clues and tests come from the measures of many $P$ and how these periods change over time.  The grand challenge question for CVs (especially the RNe and the symbiotic stars) is whether these are the progenitors of Type Ia supernovae, with this coming down to the WD masses and how they change, as well as to the detection and frequency of neon novae, as well as their population density.

For demographics, our community needs to have collective measures of nova properties, with these measures being hard-fought for each individual nova.  Historically, the first large collection of nova data was in the great book {\it The Galactic Novae} by Payne-Gaposchkin (1964).  This is still a good source of information on the old novae, especially for the detailed spectral evolution.  Duerbeck (1987) constructed a great catalog of all nova, doing nice and exhaustive work on vetting each nova and counterpart, presenting the best sky coordinates, finder charts, and collecting the primary references.  The Downes \& Shara catalog (Downes \& Shara 1993, updated to early 2006 in the on-line version\footnote{\url{https://archive.stsci.edu/prepds/cvcat/index.html}}) is a wonderful collection for all CVs of finder charts, positions, magnitude ranges, $P$, and key references.  In the last two decades, the catalog of record is the Variable Star Index\footnote{\url{https://www.aavso.org/vsx/}} (VSX) of the American Association of Variable Star Observers (AAVSO), which contains exhaustive lists for all CVs of positions, $P$, characteristic magnitudes, and many references.  The VSX is continuously up-dated from published papers, for all variable stars, as part of a titanic effort.

For the specific needs of nova demographics, these wonderful catalogs have some gaps.  All the catalogs only record the primary information (coordinates, magnitude ranges, and some timing information) while ignoring nova properties of importance for demographic issues.  Thus, none of the catalogs tabulate $M_{\rm WD}$, the light curve class, the decline rates ($t_3$ and $t_2$), the spectral class, the full-width-half-maximum (FWHM) velocity of the Balmer line profiles, the population (disk versus bulge), the distance $D$, the extinction $E(B-V)$, the derived absolute magnitudes for peak ($M_{\rm V,peak}$) and for the average quiescence ($M_q$).  Nor do these catalogs note any of the characteristic properties of importance, like for neon novae, eclipsing systems, red giant companions, $\gamma$-ray emission, nova shells, superflares, polars, intermediate polars, and asynchronous polars.  These unlisted nova properties are needed for the front-line science demographics questions.

To help with these nova demographics questions, I have been constructing comprehensive data bases for various key properties.  Strope, Schaefer, \& Henden (2010) made exhaustive compilation of the entire eruption light curves for the 93 best-observed novae, measured many light curve properties, and defined the light curve classes of S, P, O, C, J, D, and F.  Schaefer (2010) made a comprehensive set of newly measured and collected photometric properties of all ten galactic RNe, deriving many of the fundamental nova properties for all ten.  Schaefer (2022a) discovered 49 new $P$ values for novae, as well as vetting and collecting other published periods to make a total of 156 reliable periods.  Schaefer (2023b) reported on a career-long program to measure orbital period changes for 14 novae, both steady changes in quiescence ($\dot{P}$) and sudden changes across an eruption ($\Delta P$).  This program is exhaustive, in both the sense of the amount of tedious work required every year for decades and the sense that there are scant possibilities for increasing the number of measures even into the middle-term future.  Schaefer (2022b) made exhaustive measures of the distances to all 402 Galactic novae, based on the {\it Gaia} parallaxes plus all previous distance measures included as priors.  This paper collected into one big table many of the nova properties past the standard catalogs, including the spectral classes, the FWHM of emission lines, and the extinction.  Further, I went back to many original papers to measure the light curve properties of the peak magnitudes, decline rates, the light curve classes, and to derive $M_{\rm V,peak}$ for all novae.  The result is Table 6 of Schaefer (2022b), which contains all known information for all 402 Galactic novae for most of the fundamental nova properties.  And this is all collected together in one convenient table, rather than scattered over one page or one paper for each nova or property.

This Table 6 provides the easy basis for nova demographics.  Nevertheless, this table would be more useful with the addition of columns for WD mass and the quiescent brightness level.  So that is a primary goal of this paper, to provide the best possible measures of $M_{\rm WD}$ and $V_q$ for all 402 novae.  Further, these values are used to answer a variety of long-time puzzles.

\section{NOVA PROPERTIES}

This section will describe the collection and calculation for the best measures of the white dwarf mass ($M_{\rm WD}$) and the quiescent brightness level in the $V$-band ($V_q$).  Further, it will combine with the previous big table to create a larger table of the fundamental nova properties.

Previous lists of WD masses have included 92 measures derived by Shara et al. (2018), 89 model measures derived in a series of papers by M. Kato (Keio University) and I. Hachisu (University of Tokyo), and 12 measures from radial velocities in the catalog of Ritter \& Kolb (2003, with updates to 2016).  In this paper, I produce 73 new measures of $M_{\rm WD}$ using the method of Shara.  I collect a total of 293 $M_{\rm WD}$ measures.  These measures are combined together as a median for each nova with multiple measures.  The result is a list of 208 Galactic nova $M_{\rm WD}$ measures.

For two decades, the VSX catalog has served as the replacement for the moribund GCVS catalog.  This awesome and exhaustive compilation is an on-going frequently-updated product of the AAVSO that covers all variable stars.  The VSX catalog formally lists the minimum magnitudes for all Galactic novae.  For this, a researcher can query each nova one at a time, and can compile their own list of $V_q$.  A substantial problem for nova demographics with the VSX magnitudes is that they quote the faintest magnitude ever observed, including for eclipses and various low states, so the quoted minimum brightness is often not representative of the average accretion rate in quiescence in the decades near the nova eruption.  Further, for many poorly observed novae, the quoted value is often just a weak limit.  Further, the minimum brightness quoted is often for a wide variety of bands (typically $g$, $B$, $V$, and $R$).  In this paper, I collect $V_q$ measures for all 402 Galactic novae into one convenient table.  For this, I pull out average magnitudes in quiescence from many sources, including my own measures from the Harvard archival plates, from the regular AAVSO light curves, and from the AAVSO's APASS all-sky photometric survey going deep.  The result is 228 measures (not limits) for $V_q$ for Galactic novae.

These $M_{\rm WD}$ and $V_q$ measures are collated into Table 1, listing most of the fundamental properties for all 402 Galactic novae.  The base list was started with my exhaustive collection of orbital periods ($P$), including 49 new periods, that appeared in Schaefer (2022a).  This base list was expanded to include all then-known Galactic novae and many fundamental properties in Schaefer (2022b).

Table 1 contains a listing of all the fundamental properties of all 402 Galactic novae (with peaks before the middle of the year 2021).  This is printed out in full in this paper so that browsing is easy, quick lookups are possible, and this avoids sometimes-hazardous or obscure downloading.  The first column gives the standard variable star name in the standard ordering.  The second column gives the time of the peak to the nearest tenth of a year, although the recurrent novae with many eruption years are listed only as `RN'.  The third column lists the light curve class with the $t_3$ decline rate from peak in days in parentheses.  The light curve classes are defined in Strope, Schaefer, \& Henden (2010), with S-class for simple and smooth declines, P-class for smooth declines with a plateau near the transition,  PP-class is an optional variant of P for the three novae with {\it two} plateaus, O-class for novae that show oscillations or nearly-periodic flares somewhat after peak, C-class for smooth nova light curves that have a single cusp-shaped rise around the time of transition, J-class for novae that display multiple chaotic 0.5--2 mag jitters around their wide peaks, D-class for light curves that experience a dust dip after the peak, and F-class for novae that have a flat-topped light curve.  The fourth column gives the $V$-magnitude of peak, $V_{\rm peak}$.  The next column gives the average $V$-magnitudes of the nova in quiescence, $V_q$.  The next column gives the eruption amplitude in magnitudes, as calculated from $V_q$-$V_{\rm peak}$.  The 7th column quotes the spectral class (Fe II, hybrid, or He/N), `helium' for the unique helium nova V445 Pup, and `Ne' for neon novae.  The 8th column gives the FWHM of the H$\alpha$ emission line around the time of peak, as a measure of the expansion velocity of the ejecta.  The reported values are actually a mixed bag for measures throughout the time around the peak and for other hydrogen lines and for various measures of the velocity.  The next column gives the orbital period $P$ in days, as taken from Schaefer (2022a).  The tenth column gives $M_{\rm WD}$ in units of solar masses, which is one of the main products of this paper.  The 11th column gives the Galactic population (disk or bulge), from Schaefer (2022b), with capital letters indicating a high confidence assignment.  The 12th column gives the nova distance, $D$, in parsecs from Schaefer (2022b), calculated using the correct priors for novae and all previous distance measures with appropriate error bars.  These distances are strongly preferred over all prior published distances, including those calculated by the {\it Gaia} team (because they used priors not appropriate for either nova population, and they did not use the other good distance measures as appropriate priors).  In this column, the parenthetical range is for the central 68\% of the distribution from the full Bayesian analysis.  The next column gives the absolute $V$-magnitude at peak, $M_{\rm V,peak}$ in magnitudes, as calculated from $V_{\rm peak}$, $D$, and $E(B-V)$.  The next column gives the average absolute $V$-magnitude in quiescence, $M_q$ in magnitudes, as calculated from $V_q$, $D$, and $E(B-V)$.  The second-from-last column gives the time in days for the nova to fade by 2.0 mag from peak, $t_2$.  The last, 16th, column lists a variety of properties that are distinct for each nova.  These properties are notated with short acronyms, as listed in the footnote.  

Table 1 appears in the on-line electronic version with five more columns of data for all 402 novae.  These five added columns are for data that are not intrinsic to the novae, but rather are incidental due to the position of the nova in our Galaxy.  Nevertheless, each item is still needed for any of various analyses and demographics questions.  These five extra columns are placed immediately in front of the last column with various properties.  The first three of the added columns give the Galactic longitude, the Galactic latitude, and the angular distance from the Galactic center, all in units of degrees.  These are the critical properties for assigning the Galactic disk or bulge population.  The fourth added column is the excess color, $E(B-V)$ in magnitudes, which is required for removing the effects of extinction, with this being critical for estimating absolute magnitudes and luminosities.  The fifth of the added columns is the {\it Gaia} parallax in units of milli-arc-seconds, which is important for getting the best distance $D$, as listed in the 12th column.  If the fractional error is small, then the inverse of the parallax (in milli-arc-sesconds) is useably close to the best distance in units of kiloparsecs.  However, in all cases, the distance should be calculated with a Bayesian analysis that uses the priors appropriate for the nova distance distribution, with the required analysis in Schaefer (2022b).  Further, the best distances must use all previous measures with realistic error bars for additional priors, as done in Schaefer (2022b).  With these additions, the electronic table has 21 columns for 402 novae.

I have made a variety of updates and corrections from the prior list of nova properties in Schaefer (2022b).  These changes include the following examples.  For the superflaring RN V2487 Oph, Rodr\'{i}guez-Gil et al. (2023) used a radial velocity curve to discover the true $P$ to equal 0.753$\pm$0.016 days, rather than the photometric periodicity of 1.24 days from the superflares.  For V745 Sco, I have corrected the effective temperature of the red giant to the reasonable 3100 K so as to get a better estimate for $P$ of 930 days.  For V721 Sco, I have used the original eruption plates at Harvard to show that the catalogued sky position was substantially wrong, with the previous quiescent counterpart thus being wrong, so the nova is shifted from the disk population to the bulge population (Schaefer 2025b).  I have examined the original Harvard plates for EL Aql and X Ser to make an improved evaluation for their light curve classes to be C(25) and F(730) respectively.  For T CrB, the peak brightness is updated to $V_{\rm peak}$=1.7, as based on the recently recovered observations by A. S. Kamenchuk and M. Woodman (Shears 2024).  For the unique helium nova V445 Pup, I have discovered the orbital period and measured the steady period change ($\dot{P}$) for the pre-eruption and post-eruption times, as well as the sudden orbital period change ($\Delta P$) across the 2000 eruption (Schaefer 2025a).  For V659 Sct, Munari, Righetti, \& Dallaporta (2022) report nine fast flares, roughly periodic, before the transition phase\footnote{The transition phase in a nova light curve is around the time when the brightness is $\sim$3.5 mag below peak, usually when the initial fast fade switches to a relatively slow fade.  This is when the expanding shell becomes optically thin, and when the nebular emission lines appear.}, so the correct light curve classification is O(14).  I have added my newly discovered orbital periods for 12 novae (Schaefer 2025b).  I have added many novae as being shell novae (Santamar\'{i}a et al. 2025), neon novae, and as $\gamma$-ray novae.  I have not updated this listing with the novae peaking after mid-2021, nor have I yet added V407 Cyg, which was mistakenly omitted from the big list.

Each line in Table 1 shows the unique picture that describes the personality of each nova.  With the year, light curve class, $V_{\rm peak}$, $t_2$, $t_3$, and $V_q$, we can sketch a fairly accurate light curve.  With the spectral class and the FWHM, we can know the dominant emission lines and their profiles.  With $P$ and $M_{\rm WD}$, we get a good picture of the binary and the companion star.  $M_q$ and $A$ give the accretion rate for most of the nova systems.  The galactic positioning of the novae comes from $D$, the population assignment, the $E(B-V)$, the Galactic latitude and longitude, and the constellation in the star name.  And the properties column identify the various uncommon traits, like shells, eclipses, polars, and extreme superflares, that characterize each nova.  In all, each line lists all the properties for each nova that show the full picture, making for a unique and recognizable nova.

\begin{longrotatetable}
\begin{deluxetable*}{lccccccccccccccl}
\tablecaption{Properties of Galactic Novae}
\tablewidth{600pt}
\tabletypesize{\scriptsize}
\tablehead{
Nova   & Year &   LC  & $V_{\rm peak}$  & $V_q$  & $A$   &  Spec.   &  FWHM  &  $P$  &  $M_{\rm WD}$  &  Pop.  &  $D$  & $M_{\rm V,peak}$  &  $M_q$  &  $t_2$  &  Properties$^a$  \\
  &     &  class & (mag) & (mag)  &  (mag)  &  class  & (km/s)  &  (d) & ($M_{\odot}$)  &     &  (pc)  &  (mag)  &  (mag) &   (d)      &
}
\startdata
OS And	&	1986.9	&	D(23)	&	6.50$\pm$0.10 	&	17.5	&	11.0	&	...	&	1740	&	...	&	1.09	&	DISC	&	4150 (3550--5350)	&	-7.06$\pm$0.48 	&	3.9	&	11	&	Shell	\\
CI Aql	&	RN	&	P(32)	&	9.00$\pm$0.10 	&	16.1	&	7.1	&	...	&	2250	&	0.61836	&	1.20	&	DISC	&	2760 (2560--3140)	&	-5.84$\pm$0.96 	&	1.3	&	25	&	Ecl, $\Delta$P, $\dot{P}$	\\
DO Aql	&	1925.7	&	F(900)	&	8.50$\pm$0.10 	&	18.0	&	9.5	&	Fe II	&	...	&	0.16776	&	0.41	&	DISC	&	3150 (2570--4830)	&	-4.89$\pm$0.76 	&	4.6	&	295	&	Ecl, Shell	\\
EL Aql	&	1927.6	&	C(25)	&	5.5$\pm$0.2 	&	18.9	&	13.6	&	...	&	1230	&	...	&	1.04	&	DISC	&	4420 (3780--6450)	&	 -10.70$\pm$0.80 	&	2.7	&	18	&	...	\\
EY Aql	&	1926.7	&	S(40)	&	9.69$\pm$0.54 	&	19.2	&	9.5	&	...	&	...	&	...	&	1.00	&	DISC	&	5820 (4210--9610)	&	-7.54$\pm$1.87 	&	2.0	&	20	&	...	\\
V356 Aql	&	1936.7	&	J(140)	&	7.00$\pm$0.10 	&	18.3	&	11.3	&	Fe II	&	1140	&	0.42651	&	0.57	&	DISC	&	3240 (2670--5730)	&	-6.79$\pm$1.14 	&	4.5	&	127	&	...	\\
V368 Aql	&	1936.8	&	S(42)	&	5.30$\pm$0.20 	&	15.7	&	10.4	&	Fe II	&	...	&	0.69051	&	1.06	&	DISC	&	2530 (2320--2950)	&	-8.58$\pm$0.70 	&	1.8	&	15	&	Ecl	\\
V500 Aql	&	1943.7	&	S(43)	&	6.13$\pm$0.30 	&	17.3	&	11.2	&	He/N	&	1400	&	0.14526	&	1.02	&	DISC	&	3290 (2660--4740)	&	-6.95$\pm$0.70 	&	4.2	&	20	&	...	\\
\enddata
\end{deluxetable*}
\end{longrotatetable}

\subsection{White Dwarf Masses}

The WD masses hold critical demographics information for novae, so I want to compile a comprehensive list for correlation with other properties.  Many ways to measure $M_{\rm WD}$ have been used in the literature.  The standard method must be the use of radial velocity (RV) curves (Section 2.1.1), but this method is riddled with systematic problems and large uncertainties.  Alternatively, the excellent models of Hachisu \& Kato are calibrated with light curves in the optical, ultraviolet, and X-ray (Section 2.1.2).  Another excellent set of models is presented in Shara et al. (2018), where the eruption amplitude ($A$) and the decline rate (specifically $t_2$) are mapped onto $M_{\rm WD}$ and the accretion rate $\dot{M}$ (Section 2.1.3).  A variety of additional methods are reported in Section 2.1.4, including values based on the recurrence times of RNe and on X-ray properties (including the duration of the supersoft phase).  In Section 2.1.5, I analyze the numbers to get an approximate average error bars for $M_{\rm WD}$, and I go looking for systematic biases in the various methods.  In Section 2.1.6, I empirically correct one of the methods for a systematic bias, then combine all the measures as a median to form the final best $M_{\rm WD}$.

\subsubsection{Radial Velocity Measures}

The radial velocity (RV) measures of $M_{\rm WD}$ are widely viewed as the `gold standard'.  Nevertheless, for novae, these measures are poor in many ways, so that I would consider them only as a `bronze standard' or a `tin standard' or sometimes only as a `tin-foil standard'.  Bad problems include the usual large uncertainties from knowing the inclination and the companion mass, the ubiquitous problem that the published radial velocities are often poor measures of the stellar velocities, and the problem that radial velocity measures are usually greatly inconsistent with other radial velocity results and inconsistent with the astrophysical constraints.

One set of RV problems is that both the inclination and companion mass are required to get anything better than a lower limit from a mass function.  Even for the uncommon eclipsing binaries with known inclinations, the companion mass is usually only known from guesswork and analogies with main sequence stars.  The result is that most RV masses have moderately large error bars.  For example, for V1500 Cyg, Horne \& Schneider (1989) can only say that the WD mass is $\gtrsim$0.9 $M_{\odot}$, which I would represent as 1.15$\pm$0.25 $M_{\odot}$.  For GK Per, Morales-Rueda et al. (2002) can only constrain the mass to be $\ge$0.87$\pm$0.24 $M_{\odot}$, which includes masses up to the Chandrasekhar limit.  The quoted error bars on the mass usually are only allowing for RV measurement errors and the uncertainties from the inclination and the companion mass.  This makes the published error bars too small, because they do not include the real systematic uncertainties.

Gilmozzi \& Selvelli (2019) conclude ``Regrettably, $M_{\rm WD}$ is not accurately known by direct observations because of the several and severe problems one encounters in determining the primary mass and other system parameters from the observed radial velocity curves.  One example is that the velocities may not be those of the star(s), for example if they originate in or above the disk (see Wade \& Horne 1988; Thorstensen et al. 1991; Marsh \& Duck 1996).''  Wade \& Horne (1988) demonstrate that four alternative methods of analysis can return masses of 0.555, 0.836, 0.934, and 1.164 $M_{\odot}$, all for the same data for the same star.   Thorstensen et al. (1991) demonstrate that many CVs have the sinusoidal RV curve out of phase with the eclipses, by up to 76$\degr$ in the orbit, so the emission line velocities cannot represent the orbital velocity of any star.  Cantrell \& Bailyn (2007) collect data for many novae and CVs, where 75\% have significant phase offsets.  Marsh \& Duck (1996) show that the RV curves of the companion star can have systematic errors at least up to 40\% in the semi-amplitude due to effects of irradiation on the hemisphere nearest to the WD.

Nova radial velocity measures are often inconsistent.  For HR Del, the published WD masses include  1.00$\pm$0.15 (Hutchings 1979), 0.9$\pm$0.1 (Bruch 1982), 0.595$\pm$0.030 (K\"{u}rster \& Barwig 1988), and 0.60$\pm$0.10 (Selvelli \& Gilmozzi 2019), all in units of solar masses.  For CP Pup, White et al. (1993) collect five RV measures that span from 0.12--0.6 $M_{\odot}$, although the WD mass is certainly\footnote{CP Pup has a P-class light curve for a $>$0.95 $M_{\odot}$ WD, has $t_3$=8 days for 1.37 $M_{\odot}$ WD, a FWHM=2000 km/s for a mass of 1.17 $M_{\odot}$, and a He/N spectral class for $>$1.15 $M_{\odot}$.} $\gg$0.6 $M_{\odot}$.  

The published masses are often astrophysically impossible.  For CI Aql, Sahman et al. (2013) used an RV curve to measure the WD mass to be 1.00$\pm$0.14 $M_{\odot}$, whereas strong theory (e.g., Figure 7 of Shen \& Bildsten 2009) requires an RN with a 24 year recurrence time scale to be $>$1.25 $M_{\odot}$.  And we have the fateful RV result that the accreting star in T CrB is $>$2.1 $M_{\odot}$ (Kraft 1958), resulting in a decades-long red herring for the theory of RNe.  This mistaken result was compounded with Kenyon \& Garcia (1986) reporting $>$1.6 $M_{\odot}$.  For U Sco, Thoroughgood et al. (2001) make a horrible-looking claim that the WD is 1.55 $M_{\odot}$, although this is not really horrible with their quoted uncertainty of $\pm$0.24 $M_{\odot}$.  Further for U Sco, Johnston \& Kulkarni (1992) report RV masses for the WD as 0.23$\pm$0.12, 0.46$\pm$0.15, 0.29$\pm$0.14, and 0.60$\pm$0.14 $M_{\odot}$.  For KT Eri, the formal RV analysis gives 2.36 $M_{\odot}$ for the best inclination and companion mass (Schaefer et al. 2022).  For V2487 Oph, Rodr\'{i}guez-Gil et al. (2023) claimed that the companion star has a mass of 0.21 $M_{\odot}$, which is impossible given their discovery of the 0.753 day orbital period.

Nevertheless, the RV mass measures are at least a model-independent measure of the mass for 26 novae.  I have collected these from the Ritter \& Kolb catalog (2003, as updated to 2016), Pagnotta \& Schaefer (2014), Selvelli \& Gilmozzi (2019), many of the Schaefer references in the bibliography, and references therein.

\subsubsection{Model Masses From Hachisu \& Kato}

Drs. Hachisu \& Kato have a detailed model of nova light curves in the optical, ultraviolet, and X-ray bands.  Their theoretical light curves are constructed with free-free and photospheric emission, as part of a `universal decline law'.  Their models typically aim to reproduce the $U$-, $B$-, and $V$-band light curves, and the ultraviolet light curve.  They also aim to reproduce a variety of spectral data (like the start of the nebular phase and when the wind ends) and compositional data (for example to determine whether the WD is CO or ONe in composition).  Their detailed calculations include fits for the extinction, distance, and $M_{\rm WD}$.  Their typical quoted error bar is $\pm$0.06 $M_{\odot}$.

Their $M_{\rm WD}$ values have appeared in an impressive series of papers over the last 25 years.  These citations are Hachisu \& Kato (2001, 2004, 2006, 2007, 2010, 2014, 2015, 2016, 2018, 2019, 2021, 2023), Hachisu, Kato, \& Luna (2007), Hachisu, Kato, \& Schaefer (2003), Hachisu, Kato, and Walter (2025), Hachisu et al. (2002), and Kato \& Hachisu (2003).  A number of these novae have their masses updated in later papers, as slightly better input is obtained.  These come to a total of 89 novae.  I label this collection of WD masses as `H\&K' ($M_{\rm H\&K}$).

\subsubsection{Model Masses From Shara et al. (2018)}

Shara et al. (2018) construct models for generic nova light curve models, and they find a one-to-one correspondence between the input properties of $M_{\rm WD}$ and $\dot{M}$ as a function of the two properties of the eruption amplitude ($A=V_{\rm peak}-V_q$) and the time for the nova to decline by 2.0 mags from peak ($t_2$).  With this, they can derive the WD mass just from two common measures from nova light curves.  They applied their method to the collection of the 93 best observed nova light curves in Strope, Schaefer, \& Henden (2010).  This method is not applicable to any novae that does not have a normal hydrogen-rich main sequence companion star.  Further, they applied their model to the 10 known Galactic RNe, using the flash duration (taken to be the time from the first brightening to the time of the final decline) and the recurrence time scale.  This model result works even for the case with subgiant and red giant companion stars.  In all, they published 92 $M_{\rm WD}$ values for the brightest and most important novae.  This number must get lowered to 90, because subsequent research has shown two of their novae to have companion stars above the main sequence.



I have placed the Shara $M_{\rm WD}$ values on a plot of $A$ versus $t_2$ (see Figure 1), drawing contour lines of constant WD masses.  These contours are largely constant in $t_2$ within a range of 8$<$$A$$<$14 mag.  That is to say, novae away from extremes in amplitude have $M_{\rm WD}$ as a simple function of $t_2$ alone.  I have applied the method of Shara et al. to novae not in the original list.  In particular, I have placed each new nova onto the $A$ versus $t_2$ plot, interpolating to get the WD mass.  I have been careful to not include novae with subgiant or red giant companion stars.  I only applied this to novae with a reliable value for $A$ and $t_2$, all without limits.  I have not applied this to any novae significantly outside the range for which the original Shara plot is not populated by their original novae.  I have not applied the method to the unique helium nova V445 Pup.  The result is 72 new measures of the WD mass.  This brings a total of 162 masses by this one method.  I will label these WD masses as being the `Shara masses' ($M_{\rm Shara}$).

\begin{figure}
	\includegraphics[width=1.01\columnwidth]{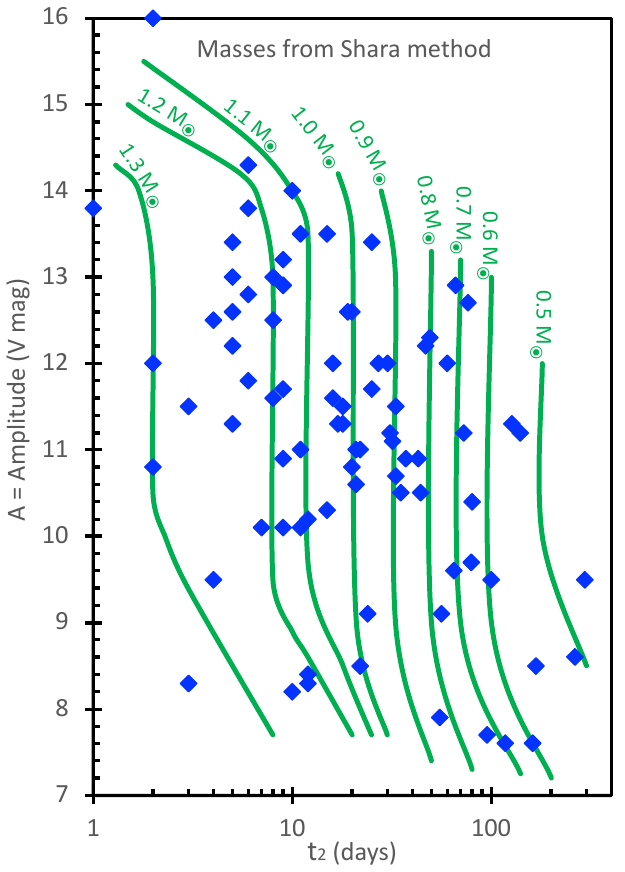}
    \caption{Shara method for measuring WD mass, with contours corrected for bias.  The method of Shara et al. (2018) is to theoretically model nova light curves as a function of the WD mass ($M_{\rm WD}$) and the accretion rate, then calculating the decline rate ($t_2$), and the nova amplitude in the $V$-band ($A$).  For any choice of $t_2$ and $A$, a unique mass is returned.  The figure plot the position of each of 80 nova analyzed in the Shara paper, with each point having a calculated mass.  I have drawn contours of equal-mass as a family of green curves, one curve for each labelled mass.  So for other novae not in the original Shara sample, a user need only plot the nova based on the observed $t_2$ and $A$, then read off $M_{\rm WD}$ from the contour lines.  This allows for the extension of the Shara model calculations to many additional novae.  A bias has been found for the smaller Shara masses, with this corrected by Equation 1, as represented by the contours in this plot.  }
\end{figure}

\subsubsection{Other Mass Measures}

For RNe, the mass can be estimated with fairly good accuracy from just the recurrence time scale, possible with the addition of measures of $\dot{M}$.  This can be done with good confidence from the Nomoto plot, like in Figure 7 of Shen \& Bildsten (2009).

The WD mass can also be estimated from X-ray properties, such as the duration of the supersoft source extending after the end of the eruption.  For the case of CP Pup, Veresvarska et al. (2024) report on WD masses from three groups using X-ray observations, yielding $>$1.1, 0.8$^{+0.19}_{-0.23}$, and 0.73$^{+0.12}_{-0.11}$ $M_{\odot}$, for which the contradictions and error bars do not inspire confidence in the method.

Various authors have estimated WD masses by combinations of methods, trying to come to some sort of a consensus value.  So Schaefer (2020) gets 6 nova $M_{\rm WD}$ values, as based on a variety of evidences including the decline rates, RV curves, and recurrence time scales.  Selvelli \& Gilmozzi (2019) estimate 18 masses from methods including decline rates, RV curves, and ultraviolet spectra.  There is no easy or reliable way to know the real error bars on these estimates, but even the internal errors suggest that the accuracy is perhaps 0.1--0.2 $M_{\odot}$.

\subsubsection{The Real Uncertainties and Systematic Biases}

The real total uncertainties in the RV masses can be estimated from two types of comparisons.  The first comparison type is where multiple RV masses are measured for one nova system, where the scatter above the reported measurement errors shows the additional unreported systematic problems.  For HR Del, with RV masses of 1.00$\pm$0.15, 0.9$\pm$0.1, 0.595$\pm$0.030, and 0.60$\pm$0.10 $M_{\odot}$, the RMS scatter is 0.21 $M_{\odot}$, so the variance above the reported error bars is 0.18 $M_{\odot}$.  For DQ Her, we have masses of 1.09 (Robinson 1976), 1.0$\pm$0.1 (Hutchings et al. 1979), and 0.60$\pm$0.07 $M_{\odot}$ (Horne et al. 1993), demonstrating that systematic errors of order 0.4 $M_{\odot}$ are common.  For V838 Her, the Ritter \& Kolb Catalog gives 0.87$\pm$0.12 $M_{\odot}$, while Garnavich et al. (2018) gives 1.38$\pm$0.13 $M_{\odot}$, so it is clear that half of these measures have systematic errors of near 0.5 $M_{\odot}$.  For CP Pup, White et al. (1993) collects RV masses of 0.12, $<$0.4, 0.6, 0.6, and $<$0.18 $M_{\odot}$, while the real mass must be $\sim$1.25 (see footnote 4), for systematic errors ranging from 0.6--1.1 $M_{\odot}$.  From these comparisons, we see that the majority of RV masses have systematic errors from 0.4--1.1 $M_{\odot}$.  This is why RV masses, at least for novae in general, are a `bronze standard' or a 'tin standard', or even a 'tin-foil standard'.

The second type of comparisons are for RNe, which provide a nice opportunity to evaluate RV masses, because the recurrence time scales prove\footnote{The strong theory reason is that the only way to get the recurrence timescale under one century is to have a high mass WD compressing the layer of accreted gas to such a high degree that the pressure at the base of the accreted layer is sufficiently high so as to reach the critical pressure to start the thermonuclear runaway with less than a century worth of accreted gas.  This simple and forced calculation has been long known, and presented in many papers.  An easy visualization of the calculation is in Figure 7 of Shen \& Bildsten (2009).  RNe must lie in the small triangular region in the upper right of the plot, where the recurrence timescale is 100 years and shorter.  This Nomoto plot shows the upper limit on the accretion rate such that we have a normal CV with no steady hydrogen burning.  This recurrent nova region extends to the left to some limit.  For all of the RNe below, other than T CrB itself, with recurrence $<$30 years, the WD mass must be $>$1.20 $M_{\odot}$, and of course it must be less than the Chandrasekhar mass.  For T CrB with 80-year recurrences, the WD mass can get as low as 1.12 $M_{\odot}$ for a case of a finely tuned  and constant accretion just below the stable hydrogen burning region.} that the WD is 1.2--1.4 $M_{\odot}$ for RNe with recurrence timescales of $<$30 years.  For T Pyx, Uthas, Knigge, \& Steeghs (2010) report the WD mass as 0.7$\pm$0.2 $M_{\odot}$, demonstrating an error of $>$0.5$\pm$0.2 $M_{\odot}$.  For CI Aql, Sahman et al. (2013) report 1.00$\pm$0.14 $M_{\odot}$, demonstrating an error of (0.2--0.4)$\pm$0.14 $M_{\odot}$.  For V2487 Oph, Rodr\'{i}guez-Gil et al. (2023) found that each Balmer line gave the K-amplitude to change by a factor of 2$\times$, so all they could do was adopt the WD mass of H\&K, but then derived an astrophysically impossible companion star mass of 0.21 $M_{\odot}$.  For U Sco, Thoroughgood et al. (2001) gives 1.55$\pm$0.24 $M_{\odot}$, while Johnston \& Kulkarni (1992) give variously 0.23, 0.46, 0.29, and 0.60 $M_{\odot}$ with quoted uncertainties up to 0.14 $M_{\odot}$.  For KT Eri, Schaefer et al. (2022) report a WD mass of 2.36$\pm$0.5 $M_{\odot}$.  For V3890 Sgr, Miko{\l}ajewska et al. (2021) give 1.35$\pm$0.13 $M_{\odot}$, with this being plausible.  For T CrB, Kraft (1958) reported $>$2.1 $M_{\odot}$, while Kenyon \& Garcia (1986) report $>$1.6 $M_{\odot}$, with both being impossible for a WD.  Further for T CrB, Planquart, Jorissen, \& Van Winckel (2025) give 1.32$\pm$0.10 $M_{\odot}$,  Belczynski and Miko{\l}ajewska (1998) give 1.20$\pm$0.20 $M_{\odot}$, while Hinkle et al. (2025) give 1.37$\pm$0.01 $M_{\odot}$, with these measures being reasonable.  For RS Oph, Zajczyk et al. (2005) and Brandi et al. (2009) have excellent RV curves, but find ``unrealistic masses'' and confused components that are not tied to any stellar motion, so no one quotes any WD mass, and instead they {\it assume} 1.2--1.4 $M_{\odot}$ for subsequent analysis.  This is an abysmal record for RV masses as a method.  Of the 17 published RV masses for RNe, 4 are larger than the Chandrasekhar mass, 6 are impossibly small at $\le$1.1 $M_{\odot}$, and three were hopelessly confused despite apparently excellent RV curves.  Only 4 out of the 17 RN RV curves produced a plausible WD mass.  That is, 76\% of the published RV WD masses for RNe are certainly bad, with systematic errors from 0.3--1.1 $M_{\odot}$.  This is why RV masses, at least for RNe, are a `bronze standard' or a 'tin standard', or even a 'tin-foil standard'.

How can the H\&K and Shara model masses be tested?  The obvious way is to compare them versus the RV masses.  This has problems, with relatively few novae with both types of mass measures, plus the fact that we cannot have any much confidence in the RV masses because the majority have huge systematic errors.

\begin{figure}
	\includegraphics[width=1.01\columnwidth]{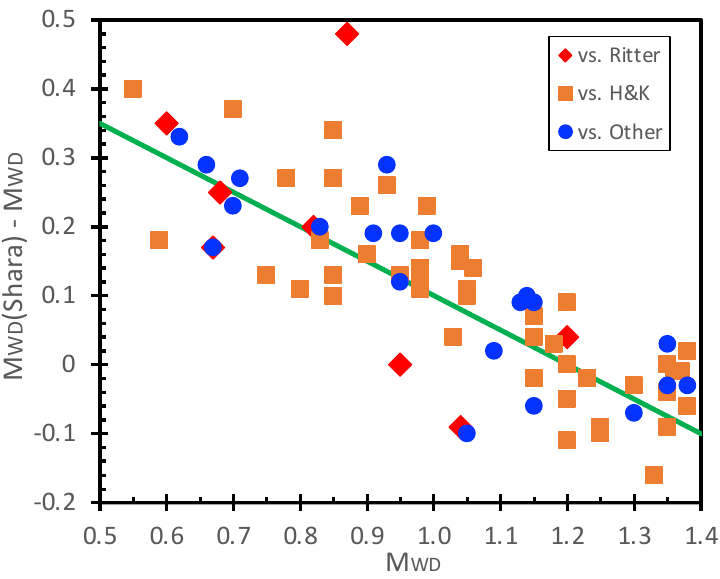}
    \caption{The differences in $M_{\rm WD}$ between the original Shara masses and masses from other sources.  This is a comparison test, where the original Shara masses vary systematically with $M_{\rm WD}$, following a linear trend (the green line).  The original Shara masses are systematically too large for low mass novae.  This is true for the three independent sets of WD masses, from the RV measures in the Ritter \& Kolb (2003) catalog, the model WD masses from Hachisu \& Kato, and the mixture of methods collected as `Other'.  The significant trend is the same for the three independent data sets, proving that the original Shara masses have a systematic bias.  This bias can be corrected with Equation 1.}
\end{figure}

Nevertheless, the H\&K masses have 7 overlaps with RV measures in the Ritter Catalog.  (I am not using the RV masses for CI Aql, V838 Her, KT Eri, U Sco, and T CrB that have obvious large errors.)  The average difference (H\&K minus Ritter) is $-$0.10 $M_{\odot}$, with an RMS scatter of 0.15 $M_{\odot}$.  This bias is not significant, and I see no correlations of significance.  If the variance of both RV and H\&K measures are comparable, then the average 1-sigma error bar for the H\&K measures is around 0.10 $M_{\odot}$.

The Shara masses (from the $A$ versus $t_2$ plot) can be compared against the 9 RV masses reported in Ritter \& Kolb (2003).  Here, there is a clear trend with mass.  The differences (Shara minus Ritter) look to be a sloped line as a function of the RV mass.  See the red diamonds in Figure 2.  For the WDs $>$0.95 $M_{\odot}$, the differences are small.  But for lower-mass WDs, the Shara method is over-estimating $M_{\rm WD}$ by 0.17--0.48 $M_{\odot}$.  The Shara et al. (2018) paper made this same test, but they only used 5 novae, with only one low-mass system.  The DQ Her WD was seen to have a 0.35 $M_{\odot}$ difference, but they made a strong case that this was due to ordinary systematic errors in the RV mass.

It is also instructive to compare the H\&K and Shara masses.  This comparison has the advantage that we have 48 novae in common, and we avoid the large systematic errors of the RV method.  The mass differences ($M_{\rm Shara}$$-$$M_{\rm H\&K}$) are plotted in Figure 2 as orange squares.  We see a highly significant and strong linear trend.  The high mass novae have only modest differences from zero, which is to say that the Shara and H\&K masses are similar above 1.1 $M_{\odot}$.  But for low-mass WDs, the Shara masses systematically deviate from the H\&K masses, being larger by 0.1--0.4 $M_{\odot}$.  The Shara et al. (2018) paper also compared the Shara and the H\&K masses.  Unfortunately, only 3 of their 37 comparisons were for low-mass WDs (in RR Pic, HR Del, and V723 Cas), and the large deviations were not noted.  Still, we are left with the case that the large systematic bias in the Shara-minus-H\&K comparisons are the same as the large and systematic bias in the Shara-minus-Ritter comparisons.

Now, alerted to the systematic problems, I can compare with one more independent data set, that being the 'Other' collection of WD masses from Section 2.1.4.  To recall, these Other measures are a collection of published masses that are a wide mixture of methods.  The 21 mass differences (Shara-minus-Other) are plotted in Figure 2 as blue circles.  Again, we see a highly significant linear trend.  This trend has near-zero differences for WDs near 1.2 $M_{\odot}$, while rising to 0.3 $M_{\odot}$ for the lowest mass WDs.  Importantly, this trend is identical to that for Shara-minus-Ritter and Shara-minus-H\&K.

We now have three independent data sets where the {\it same} trend is seen versus the Shara masses.  This triple test proves that the bias is inherent only in the Shara masses.  So there is some systematic error in the Shara method that over-estimates $M_{\rm WD}$ for low-mass WDs.  I expect that this is some sort of an ordinary problem, where some effect has an imperfect model in a complex situation.  The similar model of Livio (1991), as calibrated by Selvelli \& Gilmozzi (2019), has the same problems, where the deviations are of comparable size, yet the bias is to {\it under}-estimate the masses.  

The trend line can be represented by the green line in Figure 2.  The equation for this line is that the difference equals $0.6-0.5$$M_{\rm WD}$, with units of solar masses.  There are four problems with simply turning around a straight line.  First, there is no expectation that the model corrections are linear, so we should expect curvature.  Second, the difference equation operates off the real $M_{\rm WD}$, whereas such is not available for correcting a Shara mass by itself, so the relation should only be a function of $M_{\rm Shara}$.  The difference line can be turned around to say that the corrected-Shara mass is $2(M_{\rm Shara}-0.6)$, in units of solar masses.  Third, the simple straight line extrapolates to unreasonably low $M_{\rm WD}$, for the four novae with Shara masses under 0.8  or so.  By far the most extreme case is DO Aql, with Shara et al. (2018) reporting a mass of 0.62 $M_{\odot}$, for which the simple line would calculate a mass of 0.04 $M_{\odot}$.  So there must be a break somewhere.  DO Aql has a 244-day flat-top light curve that is more remindful of symbiotic novae, and these have WD mass frequently down to 0.4 $M_{\odot}$ (Miko{\l}ajewska 2010), suggesting that DO Aql really has a mass around that minimum.  Fourth, a similar problem arises for the highest $M_{\rm Shara}$ values, because the simple line would then push the WD over the Chandrasekhar limit.  So again, there must be some sort of a break at high mass.  All of these problems can be controlled with a correction like
\begin{eqnarray}
M_{\rm WD}  = M_{\rm Shara}~~(M_{\rm Shara}>1.2),   \nonumber	\\
M_{\rm WD}  = 2M_{\rm Shara}-1.2~~(0.9 \le M_{\rm Shara} \le 1.2),   \nonumber	\\
M_{\rm WD}  = \frac{2}{3}M_{\rm Shara}~~(M_{\rm Shara}<0.9),
\end{eqnarray}
with everything having units of solar masses.  With this correction, I find that the comparisons to the Ritter, H\&K, and Other data sets have differences with near zero averages and no trend with mass.  It is this last point that demonstrates that this correction is a reasonable approximation to some perfect theoretical model.  With this bias correction, Figure 1 shows the contours for constant $M_{\rm WD}$ as a function of $t_3$ and $A$.

I have pondered whether my big compilation of $M_{\rm WD}$ should use the masses from the published Shara method, or whether I should impose the empirical correction in equation 1.  I conclude that I must correct all the Shara masses for this bias.

This correction is of little import for most novae.  Part of the reason is that many novae have 1--3 additional measures, so the final mass as the median will usually not change greatly.  For the 110 novae for which the Shara mass is the only measure, the correction is smaller than the measurement error when the Shara mass is $>$1.1 $M_{\odot}$, so the correction will not matter much for many individual novae.  The correction will be substantial only for the 51 novae with only a Shara mass $<$1.1 $M_{\odot}$.

\subsubsection{The Final Combined $M_{\rm WD}$ Catalog}

I have 12 RV masses from Ritter, 89 light-curve-model masses from H\&K, 162 corrected $A$ versus $t_2$ masses from Shara, and 28 masses from the collection of Other estimates from mixed methods.  These measures are combined together for each individual nova as a median.  (I adopt the median, instead of the average, as it is less susceptible to outliers from large systematic problems.)  The end result is that I have $M_{\rm WD}$ for 208 novae, as tabulated in Table 1.

The uncertainty for any individual nova is hard to get with any confidence, largely because it appears that systematic errors dominate over measurement errors.  The formal published measurement errors are typically 0.05--0.20 $M_{\odot}$.  For the RV method, the strong experience from the comparisons (see above) is that the majority of RV masses have systematic errors from 0.2--1.1 $M_{\odot}$.  For the Shara method, the unaccounted-for real error bars on measuring $t_2$ can easily be $\pm$40\% (and larger for the low-mass J-class novae with jitters), leading to a systematic uncertainty in the estimated $M_{\rm WD}$ of order 0.10 $M_{\odot}$.  The difference between {\it two} measures for the same nova in Figure 2 (actually, using the values for the {\it corrected} Shara masses) has an RMS scatter of 0.14 $M_{\odot}$, so the average uncertainty in {\it one} measure is around 0.10 $M_{\odot}$.  In all, it seems that the real average uncertainty is something like 0.10--0.20 $M_{\odot}$, or $\pm$0.15 $M_{\odot}$.

The final combined catalog of WD masses (see Table 1) have the masses spanning the range from 0.41 to 1.40 $M_{\odot}$.  Five novae have $\ge$1.37 $M_{\odot}$, U Sco, V1534 Sco, V3890 Sgr, V745 Sco, and V4643 Sco.  The six novae with $\le$0.55 $M_{\odot}$ are DO Aql, V365 Cas, V1405 Cas, AR Cir, V723 Cas, and V612 Sct, of which half are neon novae.  The RNe have masses ranging from 1.20--1.39 $M_{\odot}$, with a median of 1.34 $M_{\odot}$.  The average WD Mass is 0.99 $M_{\odot}$.  The fraction of novae with $>$1.30 $M_{\odot}$ is 7\%, and the fraction of novae with $<$0.60 $M_{\odot}$ is 6\%.

\subsection{Magnitudes in Quiescence}

Nova magnitudes in quiescence ($V_q$) have a variety of problems.  Foremost is the set of problems related to the often-large intrinsic variability on all timescales for all novae.  The goal for demographics is to identify the average magnitude in some state that is representative of the long-term accretion between eruptions.  Novae in quiescence all have flickering, often with large amplitude, so for demographic purposes, we should get an average (not the faintest ever observed) magnitude from many independent times.  Novae with eclipses, often deep, should have their averages come from the out-of-eclipse data.  The V1500 Cyg class of novae are still fading, even many decades after the end of their eruptions, so we can only use the pre-eruption magnitudes.  However, novae with no pre-eruption magnitudes can be unrecognized V1500 Cyg stars with the long-term average substantially fainter than represented by the light curve even many decades after the eruption.  V445 Pup has an incredibly deep and long-lasting dust dip, which the star is still recovering from, so we can only use the pre-eruption measures.  Both RS Oph and V745 Sco have deep post-eruption dips caused by the destruction of the accretion disk by the nova eruption, so $V_q$ must be taken long after these dips are over.  For novae with both high and low states in quiescence (like T CrB), most of the accretion occurs during the high-states, so this would be the appropriate level for demographic studies.  T Pyx has been suffering a huge secular trend after its $\sim$1866 classical nova eruption (which triggered the high $\dot{M}$ and the RNe eruptions) fading in quiescence from $B$=13.8 before the 1890 RN event to $B$=16.4 currently, with the relevant magnitude for the accretion rate depending on the application, so I have simply selected the current average $V_q$.  In general, this intrinsic variability is 0.4 to 2 mag in amplitude.  This means that single measures of $V_q$ are always in disagreement with each other, often by up to 2 mags.  To get the best average $V_q$, we must average together many magnitudes on separate nights.

Another primary set of problems, for novae with faint $V_q$, is that the quiescent star is either unidentified or misidentified.  A typical case is that the nova position has an error circle of a few arc-seconds in radius, this is filled with stars in a crowded Milky Way field, and no star stands out as being the counterpart.  Sometimes, enthusiastic workers will pick one of the stars as the counterpart and this will become enshrined with scant basis for the subsequent $V_q$.  Other times, such fields will just provide a limit, which is usually not useful.  I have identified many counterparts with use of the {\it Gaia} data, and these have returned many useful measures in the {\it Gaia} pass bands (Schaefer 2022b).  The AAVSO light curve database provides $V$-band magnitudes for many obscure novae in quiescence.  VSX has collected deep limits and deep magnitudes from the observational literature, primarily from sources such as the {\it IAU Circulars} and {\it Astronomer's Telegrams}.  The great Duerbeck catalog includes limits and deep magnitudes for many faint novae.  A problem for novae with eruptions after 2005 or so is that the many of the sources only record the post-eruption fading tail of the light curves, with this not being representative of the longterm average accretion rate.  The time to fade back to the quiescent level ranges from 6 months up to decades.  The extreme case is for the V1500 Cyg stars, where these are still fading far above their pre-eruption quiescent levels even half-a-century after their eruptions (Schaefer \& Collazzi 2010).  For all the novae that are still fading in the eruption tail even today, the only solution is to recover pre-eruption magnitudes.  With all these problems, the various faint magnitudes and limits for faint novae are often inconsistent at levels of 1 mag and up.

Another set of problems for measuring $V_q$ is that the measures are often in bands other than the $V$-band, while demographic purposes require the consistency of just one band.  For this, the color of the nova is usually not known, so all that can be done is to correct as best as possible.  Such a result will necessarily have poor accuracy.  Nevertheless, we can make reasonable corrections.  For converting from $B_q$ to $V_q$, I have compiled a list of $B-V$ for novae at minimum, and then extinction corrected this with the independent measures of $E(B-V)$.  For this analysis, I have only used novae with well-measured $V_q$ and with ordinary main-sequence companion stars.  I find that the median intrinsic color is close to $B-V$=0.0, although with an RMS scatter of 0.4 mag.  So to modest accuracy, $V_q$=$B_q$-$E(B-V)$.  The only other band that frequently needs conversion is from the {\it Gaia} $b$ band (also labelled as $BP$).  This band has an effective wavelength of 5320~\AA, so it is already close to the $V$-band.  Jordi et al. (2010, Table 3) give the conversion as $V_q$$\cong$$b_q$-0.4$(B-V)$.  From a large compilation of $b_q$-$V_q$ with extinction corrections, I find that this color is near 0.0 mag, although with an RMS of 0.4 mag.  So for only modest accuracy, the correction is $V_q$=$b_q$-0.4$E(B-V)$.  These color corrections add to the uncertainties for many $V_q$ measures.

A deep problem for nova demographics is that the magnitudes in quiescence, as observed in the decades before and after the nova eruption, might be systematically different from the long-term average over the complete eruption-to-eruption cycle.  That is, it might be that the brightness far in time from any eruption is different from the brightness in the decades close to the time of the eruption.  If such a difference is substantial, then the use of my tabulated $V_q$ measures could be systematically misleading as a proxy for the average accretion rate.

In all, the measures of $V_q$ have substantial uncertainties from a variety of causes.  These error bars vary substantially from nova to nova, largely in ways that cannot be quantified.  For the well-observed novae with only modest variability, the error bar on the average is something like 0.2 mag.  For faint novae, the error bars might be $\pm$1.0 mag or worse.

My students and I have produced average $V_q$ for many novae.  For the 93 best observed novae, the average quiescent magnitudes have been evaluated and presented in Strope, Schaefer, \& Henden (2010).  For the 10 Galactic RNe, Schaefer (2010) averaged $V_q$ from large amounts of my own photometry.  For the V1500 Cyg stars that are still fading to their quiescent level, Schaefer \& Collazzi (2010) used pre-eruption magnitudes to set $V_q$.  Collazzi et al. (2009) measured and collected pre-eruption magnitudes in quiescence for 36 novae, plus I have many more pre-eruption measures from the Harvard plates.  Pagnotta \& Schaefer (2014) have curated a set of 112 $V_q$ measures, with this covering averages from many literature sources.  Schaefer (2022b) has used the {\it Gaia} data to positively identify many faint counterparts, and to list 215 magnitudes in the $b$, $g$, and $r$ bands.  For single novae of importance, I adopt the values published by Salazar et al. (2017) for V1017 Sgr, by Schaefer (2023a) for T CrB in its high state, by Schaefer (2023b) for T Pyx in recent years, by Schaefer (2025a) for V445 Pup before its eruption, by Schaefer et al. (2022) for the century-long average of KT Eri, by Schaefer (2024) for V745 Sco, by Schaefer (2025b) for V707 Sco, and by Schaefer et al. (2019) for QZ Aur.

For other novae, I have pulled $V_q$ measures from a variety of sources.  The wonderful SMARTS atlas\footnote{\url{https://www.astro.sunysb.edu/fwalter/SMARTS/NovaAtlas/atlas.html}} (Walter et al. 2012) contains many long series of photometry and spectroscopy for 113 novae, many with $V$-band light curves long into quiescence.  The Duerbeck catalog provided many magnitudes for old novae.  VSX provided minimum magnitudes for most novae, and for many novae this is the only useable information.  The AAVSO light curves provide good $V$ measures for many novae that are brighter than 17th mag or so in quiescence.  The APASS survey of the AAVSO provides $V$ measures on a handful of nights around 2010.  

I have judiciously combined all the measures for each nova.  These are now listed in Table 1.  The eruption amplitude is then calculated as $A$=$V_q$-$V_{\rm peak}$.  Further, the absolute magnitude in quiescence is calculated as $M_q$=$V_q-3.1E(B-V)+5-5\log(D)$.  For later work, these measures will be turned into average accretion rates ($\langle \dot{M} \rangle$) for demographic purposes.

\section{CORRELATIONS WITH $M_{\rm WD}$}

A part of my original reason to go through the task of deriving as many good $M_{\rm WD}$ values as possible was to use the catalog as a test of whether CV WD masses are increasing or decreasing over evolutionary time.  This task is reserved for a later paper.  The other part of my original reason to create an exhaustive catalog of nova WD masses is to test various correlations seen by myself and others.  For example, the J and D events are characterized as arising from low-mass WDs, while S and P events appear to need high-mass WDs.  And the velocity of the nova ejecta is likened to the WD mass, where high FWHM is taken to require high-$M_{\rm WD}$.  Now, with the exhaustive $M_{\rm WD}$ catalog, these tasks can be generalized to seeking correlations with many nova properties, with the aim of understanding the origin of the observed nova properties.

\subsection{Light Curve classes; S, P, O, C, J, D, and F}

\begin{figure}
	\includegraphics[width=1.01\columnwidth]{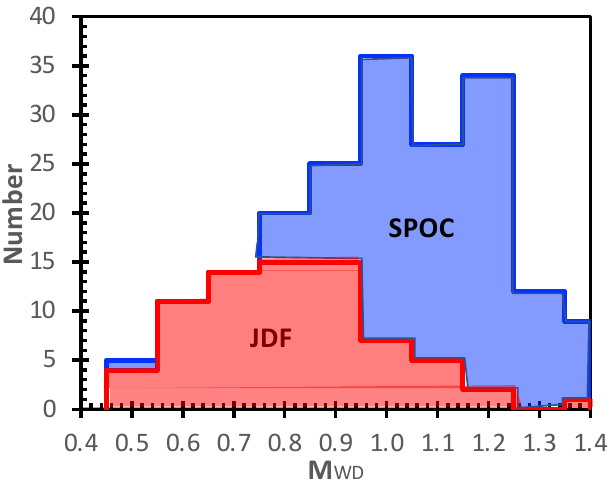}
    \caption{Histogram of WD masses for SPOC and JDF light curve groups.  The JDF light curve classes (the red shaded histogram) have a distinctly low-mass distribution, with an average of 0.81 $M_{\odot}$ and a 1-sigma range extending up to 0.95 $M_{\odot}$.  The SPOC light curve classes (the blue shaded histogram) have a distinctly high-mass distribution, with an average of 1.10 $M_{\odot}$ and a 1-sigma range extending down to 0.95 $M_{\odot}$.  The distribution is like that expected for a sharp cutoff at 0.97 $M_{\odot}$, where the known measurement errors of $\pm$0.15 $M_{\odot}$ create smearing that mixes up the two distributions around the cutoff.  }
\end{figure}

\begin{deluxetable}{cccc}
\tablenum{2}
\tablecaption{SPOC are high-mass, JDF are low-mass}
\tablehead{
\colhead{Light curve} & \colhead{Number} & \colhead{$\langle M_{\rm WD} \rangle$} & \colhead{Central 68\%} \\
\colhead{class} & \colhead{of novae} & \colhead{($M_{\odot}$)} & \colhead{range ($M_{\odot}$)}
 }
\startdata
S	&	78	&	1.09	&	0.95--1.23	\\
P	&	31	&	1.13	&	0.96--1.33	\\
O	&	6	&	1.08	&	0.99--1.14	\\
C	&	4	&	1.13	&	1.03--1.24	\\
J	&	40	&	0.77	&	0.59--0.92	\\
D	&	32	&	0.86	&	0.68--1.00	\\
F	&	4	&	0.80	&	0.48--1.11	\\
SPOC	&	119	&	1.10	&	0.95--1.26	\\
JDF	&	76	&	0.81	&	0.60--0.98	\\
\enddata	
\end{deluxetable}

The {\it shapes} of nova light curves show a wide variety, but I have defined 7 classes (S, P, O, C, J, D, and F) that still cover all events (Strope, Schaefer, \& Henden 2010).  The basic shape for all nova light curve is the simple fast rise followed by a smooth decline that transitions to a slow decline at the transition epoch, with 38\% of all novae in this S light curve class.  All other nova light curves are just variations on the S class.  The P class has a relatively flat plateau near the transition, the O class has quasi-periodic oscillations around the time of transition, and the C class has the light curve rising to a cusp-shaped maximum around the transition.  The J class novae all have several 3--20 days duration flares added on top of the basic S class light curve.  The D class novae feature a dip in brightness around the transition caused by dust formation (then dispersal) superposed on the S class shape.  The F class novae are just the S shape where the peak is a long flat maximum.  The basic S shape is well understood by the usual 1-dimensional models.  The P plateaus come from the quasi-static envelope around the WD.  The D dust dips are well known from models of dust formation.  

To try to get some understanding on the physics of light curves, I can now test to see how the light curve shape depends on $M_{\rm WD}$.  Table 2 lists the average WD mass and the central 68\% range for the seven light curve classes.  For the 195 novae in the seven classes, we see a distinctly bimodal distribution.  All four classes of S, P, O, and C have effectively identical distribution with mean masses near 1.10 $M_{\odot}$, while the three classes of J, D, and F have effectively identical distributions with mean masses near 0.81 $M_{\odot}$.  That is, S, P, O, and C classes are indistinguishable by means of masses, whilst J, D, and F classes are separate and indistinguishable.  For purposes of this section, we can lump together the four high-mass classes into a larger class labeled SPOC, and we can lump together the three low-mass classes into a larger class labeled JDF.

The binned histogram for SPOC and JDF is presented in Figure 3.  The take-away from this figure is that SPOC novae are high-mass systems, while the JDF novae are low-mass systems.  The break point is near 0.95 $M_{\odot}$.  So if we consider any single SPOC nova, then it is likely to have $M_{\rm WD}$$>$0.95 $M_{\odot}$.  Any given JDF nova is likely to have $M_{\rm WD}$$<$0.95 $M_{\odot}$.  Similarly, if a nova WD mass is high, say $>$1.2 $M_{\odot}$, then it is very likely to have a light curve shape like S, P, O, or C.  And if a nova WD mass is low, say $<$0.75 $M_{\odot}$, then it is very likely to have a light curve shape like J, D, or F.  That is, the WD mass is the primary determinant of the light curve shape.

The division at 0.95 $M_{\odot}$ is not perfect.  The most blatant deviations are that V365 Car is an S(530) nova with $M_{\rm WD}$=0.51 $M_{\odot}$, and V1186 Sco is a J(62) with $M_{\rm WD}$=1.24 $M_{\odot}$.  (V445 Pup is a D(240) nova with 1.35 $M_{\odot}$, but as the unique helium nova with greatly different explosion physics, there is no reason to think that it should follow any rule for hydrogen novae.)  V365 Car appears to be a real exception, because the light curve is certainly an S class event (Henize \& Liller 1975), while the very large $t_3$ requires a very low WD mass.  V1186 Sco has a clear J class light curve and a confident $t_2$=12 days, so it appears to be a far outlier to the 0.95 $M_{\odot}$ cutoff.  However, V1186 Sco has FWHM=500 km s$^{-1}$, which strongly points to a low-mass WD.  With contradictory indications as to high-or-low mass, this nova cannot be a confident exception to the division.

The 0.95 $M_{\odot}$ division between SPOC and JDF is not well measured by the histogram in Figure 3, because the WD masses are poorly measured, with typical real error bars of $\pm$0.15 $M_{\odot}$.  So if the underlying distribution for SPOC has a sharp cutoff at 0.95 $M_{\odot}$, then we expect that our observed histogram will have some sort of a half-Gaussian distribution extending down to 0.8 $M_{\odot}$ or so.  Similarly, a real sharp cutoff for the JDF novae will appear in Figure 3 with a half-Gaussian distribution extending from 0.95 to 1.2 $M_{\odot}$.  So the histogram in Figure 3 is consistent and expected for a sharp cut between SPOC and JDF at 0.95 $M_{\odot}$.

The division between SPOC and JDF appears to be entirely governed by $M_{\rm WD}$.  But there is no difference in mass within these two groups.  So there must be some additional parameter that distinguished S from P from O and from C classes.  And there must be some parameter past $M_{\rm WD}$ that governs whether the nova light curve has J, D, or F shape.  For this, looking in Table 1, I cannot significantly distinguish any of the SPOC classes from the others on the basis of the orbital period, the FWHM, the spectral type, the decline rates, the amplitudes, or the absolute magnitudes.  Similarly, I do not see how to distinguish between the J, D, and F classes.  So apparently the physical property that determines the choice between S-P-O-C and J-D-F is not readily visible.  The obvious idea for a nearly-invisible property of the binary that could determine brightenings and dimmings of the basic light curve is the WD magnetic field.

\subsection{Decline Rates $t_3$}

\begin{figure}
	\includegraphics[width=1.01\columnwidth]{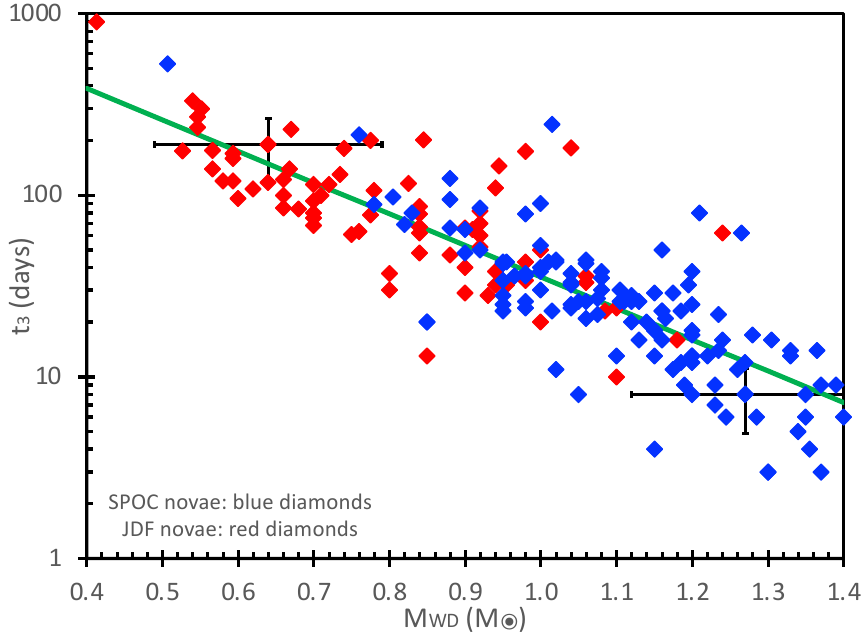}
    \caption{Decline rate versus WD mass.  From Table 1, the decline rate ($t_3$) is plotted versus $M_{\rm WD}$ for 74 JDF novae (red diamonds) and for 117 SPOC novae (blue diamonds).  We see a highly significant straight line, on this log-linear plot, with the best fit shown with the green line, with $t_3$ equaling $10^{(-1.73M_{\rm WD})}$$\times$1900 days.  The typical measurement errors ($\pm$0.15 $M_{\odot}$ in mass and $\pm$40\% in $t_3$) are displayed for two points.  The plot has substantial scatter about the green line, but all the scatter is consistent with ordinary measurement errors, so the real underlying relation apparently has only small intrinsic scatter.  }
\end{figure}

Nova light curves are defined by their timescales (quantified as $t_3$ or $t_2$) as well as by their shape.  The speed of decline has long been recognized as being a function of the WD mass (e.g., Pagnotta \& Schaefer 2014), with slow declining novae being low-mass and the highest mass WDs (e.g., for the RNe) always being short in peak durations.  Shara et al. (2018) used a detailed model, with the results that $M_{\rm WD}$ is predominantly a function of $t_2$.  Livio (1992) developed an analytic model deriving $t_3$ from just $M_{\rm WD}$, while Selvelli \& Gilmozzi (2019) provided a calibration for this equation.

With my catalog of $M_{\rm WD}$ and $t_3$, I can quantify the empirical relation.  Figure 4 is a simple plot of the logarithm of $t_3$ versus the WD mass.  We see a highly significant linear relation, in this log-linear plot, with no apparent curvature.  The best fit line is
\begin{equation}
t_3=10^{(-1.73M_{\rm WD})} \times1900~ {\rm days}.
\end{equation}
This straight line is good for both SPOC and JDF novae.  (V445 Pup was not included in this fit, because the eruption physics of a helium nova is substantially different from that of the usual classical novae.)  

The plot in Figure 4 has substantial scatter about the best fit line.  The typical error bars ($\pm$0.15 $M_{\odot}$ in mass and $\pm$40\% in $t_3$) are displayed for two points in Figure 4.  This scatter is fully consistent with the typical measurement errors around a perfect relation.  With this, it looks like the exact relation in equation 2 (i.e., the green line in Figure 4) is of good accuracy.

The WD mass is the dominant determinant for the light curve shape and is the only determinant of the light curve time scale.  That is, all nova light curves are largely dictated by $M_{\rm WD}$.

\subsection{Spectral Classes; Fe II, Hybrid, and He/N}

The optical spectra around the peak are now traditionally divided into two classes, called `Fe II' and `He/N' (Williams 1992, 2012).  The Fe II novae have the brightest non-Balmer lines being from many Fe II lines, plus other low-excitation lines, often with P-Cygni absorption components.  The He/N novae have the brightest non-Balmer lines being from helium and nitrogen high-excitation lines, with broad often-square line profiles.  Sometimes, one nova will transition from an Fe II spectrum to a He/N spectrum, with these novae being termed to have a `hybrid' spectral class.  Perhaps many or all novae actually are hybrid, but in practice one phase or another dominates, so reports in the literature usually identify particular novae as being either Fe II or He/N.  Williams (2012) states a paradigm that the He/N novae are associated with the more massive WDs.

This can be quantified and tested with my listings of $M_{\rm WD}$ and spectral classes.  In all, I have 152 novae with both properties measured.  These consist of 116 Fe II novae, 14 hybrid novae, and 22 He/N novae.  The distributions with WD masses (in units of $M_{\odot}$) is presented in Table 3, with each cell giving the number of observed novae in the given mass range and the given spectral class.  We see that novae with $M_{\rm WD}$$<$1.15 $M_{\odot}$ are mostly Fe II events.  And we see that novae with $M_{\rm WD}$$>$1.25 $M_{\odot}$ are mostly He/N events, with a minority of hybrids.

\begin{table}
	\tablenum{3}
	\centering
	\caption{Spectral Class Versus WD Mass}
	\begin{tabular}{lrrrr}
		\hline
			&	Fe II  & Hybrid & He/N & Neon \\
		\hline
Count	&	116	&	14	&	22	&	37	\\
$\langle M_{\rm WD} \rangle$	&	0.93	&	1.11	&	1.19	&	1.02	\\
68\% Range 	&	0.70--1.15	&	0.95--1.28	&	1.02--1.34	&	0.65--1.22	\\
0.40--0.45 	&	1	&	0	&	0	&	0	\\
0.45--0.55 	&	2	&	1	&	0	&	3	\\
0.55--0.65 	&	8	&	1	&	0	&	2	\\
0.65--0.75 	&	11	&	0	&	0	&	3	\\
0.75--0.85 	&	16	&	0	&	0	&	1	\\
0.85--0.95 	&	15	&	2	&	3	&	2	\\
0.95--1.05 	&	23	&	2	&	3	&	4	\\
1.05--1.15 	&	20	&	0	&	1	&	7	\\
1.15--1.25 	&	19	&	5	&	6	&	7	\\
1.25--1.35 	&	0	&	3	&	5	&	5	\\
1.35--1.40 	&	0	&	1	&	4	&	3	\\
		\hline
	\end{tabular}
\end{table}

A substantial problem with this table is that all the individual $M_{\rm WD}$ measures have a real uncertainty of $\pm$0.15 $M_{\odot}$ or so.  This makes for mixing in the vertical direction in the table.  For example, I suspect that the one He/N nova with the quoted mass of 0.85 $M_{\odot}$ (V1535 Sco) is actually of much higher mass\footnote{The light curve is P class, $t_3$ is 20 days, and the FWHM is 2000 km s$^{-1}$, with all three indicators pointing to high mass.  The $t_2$ value equals 14 days, so Shara's method yields a mass of 1.16 $M_{\odot}$.  But I have not used this Shara mass because V1535 Sco has a red giant companion star, even though this has little effect on the derived mass.  So the only official mass measure that I have is from H\&K, with $M_{\rm WD}$=0.85 $M_{\odot}$, despite numerous indicators that the mass is actually relatively high.}.  Further, for the novae that are really hybrid, then the assignment to Fe II or He/N will depend on the vagaries of when the spectral class assignment was made.  In all, the numbers in Table 3 do not have high fidelity.

We can make a crude attempt to deconvolve the numbers in Table 3 to account for the measurement errors and spot the real underlying distribution.  For the Fe II spectral class, the underlying distribution is consistent with all Fe II novae having masses $<$1.15 $M_{\odot}$ or so.  The 19 Fe II novae with masses from 1.16--1.24 $M_{\odot}$ are simply the Gaussian tail for measurement errors of WDs $\lesssim$1.15 $M_{\odot}$.  For the He/N spectral class, the underlying distribution is consistent with all having masses $>$1.15 $M_{\odot}$ or so.  The 7 He/N novae with masses $<$1.15 $M_{\odot}$ are just the expected Gaussian tail for measurement errors of WDs with $\gtrsim$1.15 $M_{\odot}$.  For the hybrid spectral class, the underlying distribution is consistent with all having WD masses $\sim$1.15 $M_{\odot}$.  The 15 hybrid novae have an average mass of 1.11 $M_{\odot}$, with an RMS scatter of 0.23 $M_{\odot}$.

So the general idea that the spectral class is predominantly determined by $M_{\rm WD}$ is confirmed, with the Fe II novae coming from low-mass WDs, the hybrid novae coming from middle-mass WDs, and the He/N novae coming from high-mass WDs.  In particular, it appears that the dividing line is fairly sharp at around 1.15 $M_{\odot}$.

Williams (2012) explains the spectral classes as being predominantly determined by the binary mass ratio, $q$=$M_{\rm comp}$/$M_{\rm WD}$.  This can be tested.  For stars with $P$$<$0.6 days or so, $M_{\rm comp}$ is well determined to be 2.6$P$ in solar masses (Frank et al. 2002, equation 4.11).  With this, I can calculate $q$ for 42 Fe II novae and 7 He/N novae.  For the Fe II novae, the average and median $q$ values are 0.62 and 0.56 $M_{\odot}$.  For the He/N novae, the average and median $q$ values are 0.61 and 0.54 $M_{\odot}$.  There is no difference in the mass ratio for the two spectral classes.  

\subsection{Neon Novae}

Neon novae are otherwise ordinary classical nova eruptions that happen to have a startlingly high abundance of neon in the ejecta.  This critical anomaly can only be detected by looking late in the tail of the eruption, in the nebular phase, with high neon abundances shown only in the strengths of neon emission lines.  In most cases, the [Ne III] lines at 3869 and 3968~\AA~in the near ultraviolet are used.  The class of neon nova was first discovered by McLaughlin (1944), with his prototypes including GK Per, RS Oph, and DQ Her.  Neon novae were rediscovered in the 1980s when detailed abundance calculations showed nova ejecta had high neon and aluminum content (Truran \& Livio 1986).  For ejecta that have more than 10$\times$ solar of neon, the only way to get such bulk quantities is from dredge-up of mass from an underlying ONe WD.  The fact that dredge-up is required demonstrates that the neon nova have more mass ejected than mass accreted between eruptions, so $M_{\rm WD}$ is decreasing over each eruption cycle, and the system cannot be a Type Ia supernova progenitor.  The fact that the WD is of ONe composition also means that there is not enough thermonuclear energy available to explode as a normal Type Ia supernova, so again neon novae cannot be progenitors.

Roughly one-third of novae are neon novae, but only a fraction of novae have been tested, so my catalog has recognized only 46 cases.  The WDs in neon nova systems must have been born from $\sim$8--10 $M_{\odot}$ stars and born with a mass of $\gtrsim$1.2 $M_{\odot}$.  This limit is critical because substantial numbers of neon nova are {\it below} this limit (e.g., Shara 1994; Takeda et al. 2018).  Such is only possible if the WD has lost large masses over each eruption cycle.  The implication is that WDs in neon novae are losing mass and such are not Type Ia supernova progenitors.  

Neon novae are indistinguishable from non-neon novae in all their basic properties, so this implies that the eruptions of non-neon novae are also ejecting large amounts of mass each eruption (relative to the mass accreted between eruptions).  This is a good argument that all novae, and hence all CVs, have their $M_{\rm WD}$ {\it decreasing} over evolutionary time scales.

Now, with my new list of $M_{\rm WD}$ and neon nova, we can look at the statistics and demographics (see Table 3), going past the statements from one or a few novae.  I have 37 neon novae with measured WD masses.  Of these, 70\% have $M_{\rm WD}$$<$1.2 $M_{\odot}$.  If we push to a limit of 1.0 $M_{\odot}$ (to allow for uncertainties in the threshold and for uncertainties in measuring the WD masses), we still have 35\% of neon nova far below the threshold.  With 13 out of 37, this is not small number statistics.  This is emphasized by one of the original prototypes (DQ Her) having a WD mass of 0.66 $M_{\odot}$.  The DQ Her WD has already lost half of its birth mass, so it must be a relatively old system.  What we see is that the majority of neon novae have evolved far past their birth condition, losing from 0.1--0.6 $M_{\odot}$ to date.  That is, neon novae are losing massive shells of the underlying ONe WD each eruption, and are being steadily whittled down in size.

The collection of 46 neon novae has the same properties ($t_3$, light curve class, spectral class, FWHM, and so on) as the general population of Galactic novae.  This points to the neon nova as having the same eruption properties as the non-neon novae, and hence for all cataclysmic variable.  That is, the existence and majority of neon novae with $M_{\rm WD}$$<$1.2 $M_{\odot}$ points to the general case of novae ejecting more mass than is accreted between eruptions.

\subsection{FWHM of Emission Line Widths}

\begin{figure}
	\includegraphics[width=1.01\columnwidth]{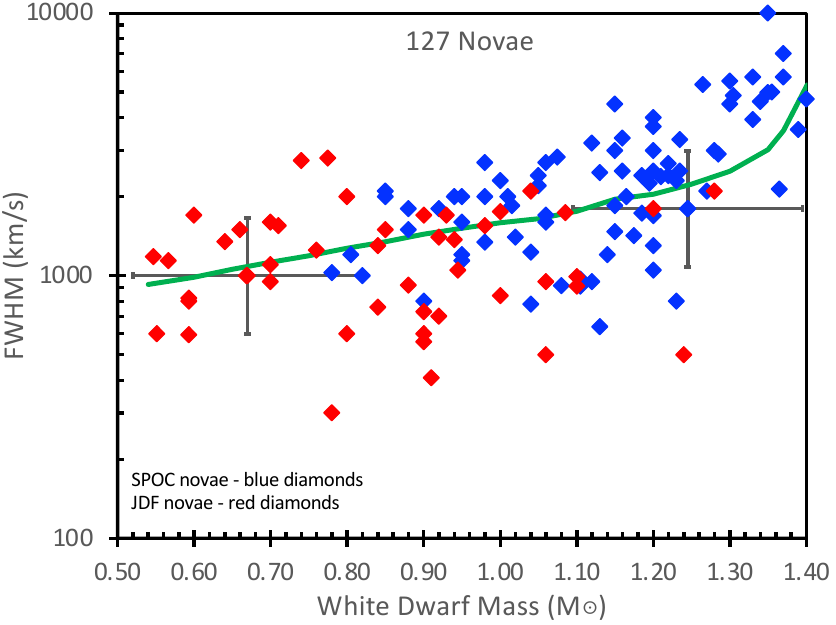}
    \caption{The FWHM is a function of $M_{\rm WD}$.  The 127 novae follow a highly significant trend.  The green curve is for 0.23 times the escape velocity of the WD.  Through most of the range, it is adequate to quantify the model as $10^{(M_{\rm WD}/2)}$$\times$500 km s$^{-1}$.  The escape velocity has an increasing deviation as the Chandrasekhar mass is approached, with this being seen both in the escape-velocity model and in the nova data.  Indeed, for masses $>$1.3 $M_{\odot}$, the observed upward deviations are substantially larger than predicted by the escape velocity alone.  The scatter is large, but is consistent with the $\pm$0.15 $M_{\odot}$ for the real uncertainty in the masses and with the $\pm$0.22 for the real uncertainty in the logarithm of FWHM.  Two typical error bars are presented for two of the points.  That is, the observations look to closely follow a single function, while the known measurement uncertainties produce the observed scatter in the plot.  With this, the real intrinsic relation is consistent with Equation 4 with no variation.}
\end{figure}

The FWHM of emission lines during the eruption is an imperfect measure of the expansion velocity of the ejecta ($V_{\rm ejecta}$).  A definitional problem is that the gases are ejected with a wide range of velocities, with no particular velocity having any meaning or utility for either theory or observation.  For example, does the FWHM correspond to some maximum velocity, some mass-averaged value for the entire depth of the shell, some average velocity above the photosphere, or half of that average?  And the observed FWHM changes by up to a factor of 2$\times$ over the eruption duration, and we have no knowledge of any time to choose as being best.  Further, nova ejecta are greatly asymmetric, so the observed FWHM arises from the happenstance of whether high-or-low velocity gas happens to be along the line of sight to Earth.   We have no way of knowing what our FWHM corresponds to in the complex environs of the ejecta, much less any standard way to compare from nova-to-nova.  In addition to these definitional problems, the measure of the FWHM has severe practical problems.  Some of these problems come because published reports only give the HWZI (Half-Width-Zero-Intensity) or the Gaussian sigma of the emission line profiles, with only variable and unknown relation to the FWHM.  Further problems arise due to large changes in profile shapes, for example due to P-Cygni dips, square-topped lines,  and multiple line components, where the formally defined FWHM will have differences by perhaps a factor-of-two from any consistent velocity.  Further, the FWHM varies with the line chosen, even amongst the hydrogen emission lines, where often the only available measure is for H$\beta$ or the Paschen series in the infrared.  Further, the line widths change greatly over time, we have no way of knowing what times to choose to standardize measures, and usually the published data quote the FWHM for a wide range of times from before the peak to after the transition.  This is all to say that my list of FWHM has severe problems with measurement errors and with definitional inconsistencies.  The various problems described likely combine to make for differences of up to a factor of two or so.  The overall average uncertainty can be estimated by looking at the vertical scatter in Figure 5, where the chi-square fit shows an average scatter of 0.22 in the logarithm of FWHM.  This suggests that the real total uncertainty in producing an average FWHM covers a one-sigma range from 0.60--1.66 times the FWHM.

Table 1 has 127 novae with both FWHM and $M_{\rm WD}$.  These are plotted in Figure 5.  We see a highly significant trend with much scatter.  For masses $<$1.3 $M_{\odot}$, the trend is fit by the line
\begin{equation}
FWHM = 10^{(M_{\rm WD}/2)}500 \rm{~km ~s}^{-1}.
\end{equation}
This relation is coincident with the straight line portion of the green curve for $<$1.3 $M_{\odot}$ in Figure 5.  The scatter is large.  The scatter is consistent with the measurement uncertainty for the WD mass of $\pm$0.15 $M_{\odot}$ and for the FWHM of $\sim$50\%.  So the underlying function could well be a tight relation.

For many classes of eruptions and outflows, the ejection velocity is comparable to the surface escape velocity of the star.  The FWHM will presumably be some constant (with little imagination, I'll label this as `C') times some characteristic ejection velocity.  So for nova ejections, the default model is FWHM/C$\sim$$V_{\rm ejecta}$$\sim$$V_{\rm escape}$$\sim$$\sqrt{2GM_{\rm WD}/R_{\rm WD}}$.  With the WD radius, $R_{\rm WD}$, being a function only of the WD mass, the FWHM should be a simple function of the WD mass.  This can be tested with my list of nova properties.

So, a reasonable model is that the FWHM is as some constant factor times the surface escape velocity of a WD of mass $M_{\rm WD}$.  For this, I have used the relativistic model for the WD radius.  With this, I have fit the data in Figure 5 to the scaled escape velocity as a function solely on the WD mass.  I find that the best fit scale factor is $C$=0.23.  So for most novae, the FWHM is nearly one-quarter the WD surface escape velocity.  With this, 
\begin{equation}
FWHM = 0.23\times V_{\rm escape} = 0.23\times \sqrt{2GM_{\rm WD}/R_{\rm WD}}.
\end{equation}
This model is displayed in Figure 5 as the green curve.  Despite the substantial scatter from measurement errors, this simple model does an excellent job of representing the 127 novae.  This even includes the upward curvature as the Chandrasekhar limit is approached.  This good agreement between the model and the averaged FWHM data suggest that the ejection velocity is simply related to the escape velocity.  

With the $C$=0.23 result, it is as if the relevant escape velocity is for an `altitude' in the nova envelope corresponding to 19$R_{\rm WD}$.  For a 1.0 $M_{\odot}$ WD (with radius 5570 km), this critical altitude is near $10^{10}$ cm or 0.15 $R_{\odot}$.  This altitude is roughly 10\% of the semimajor axis for a nova near the period gap.  With this position being so far inside the orbit of the companion star, this suggests that the ejection does not involve the companion star.

\subsection{Recurrent Novae}

Recurrent novae (RNe) are classical nova that are defined to have a recurrence time scale $<$100 years (Schaefer 2010).  To get this high frequency, the system must have {\it both} a high accretion rate ($\dot{M}$) and a near-Chandrasekhar mass WD.  From the theory calculations originally by Nomoto (1982), the RNe must have $M_{\rm WD}$$>$1.10 $M_{\odot}$, while the observed RNe have recurrence times such that $M_{\rm WD}$$>$1.20 $M_{\odot}$ (Shen \& Bildsten 2009).  There is no way around this limit.  So for RNe with recurrence times around 30-years or faster, we have the strong constraint that $M_{\rm WD}$$>$1.20 $M_{\odot}$.

RNe eruptions are always fast, and usually faint, so that most of their eruptions would have been missed over the last century.  This means that my list of all known Galactic novae must contain many true RN systems for which only one eruption happens to have been discovered.  Over the past two decades, colleagues and myself have made a wide variety of searches aimed to discover prior lost eruptions of known galactic novae (e.g., Pagnotta \& Schaefer 2014).  Up until recently, all suggestions for RN candidates have used proxy indicators for $M_{\rm WD}$, including the $t_3$, the light curve class, and FWHM of emission lines.  Now, with my long list of $M_{\rm WD}$, we can directly test whether the WD is massive enough to support a fast recurrence time scale.

A further realization is that RN candidates must {\it also} have high-$\dot{M}$, certainly $>$2$\times$10$^{-8}$ $M_{\odot}$ yr$^{-1}$ (Shen \& Bildsten 2009).  The accretion rate can be derived either from the quiescent absolute magnitude ($M_q$) or with the method of Shara et al. (2018) that keys off the eruption amplitude.  With this, some of the best old RN candidates are seen to have no chance for a fast recurrence time because their accretion rate is too low by far.  One of the most hopeful old candidates was V838 Her, for which many proxy indicators showed an extreme high mass, with P(4) light curve class, He/N spectral class, and a FWHM of 5000 km s$^{-1}$.  Now, various investigators give $M_{\rm WD}$ as 1.36, 1.35, and 1.38 $M_{\odot}$, so V838 Her surely satisfies half of the RN requirement.  But V838 Her has its accretion disk rather faint at $M_q$=5.1, while Shara et al. (2018) calculates that $\log$[$\dot{M}]$ (in $M_{\odot}$ yr$^{-1}$) is -10.02, with a recurrence time scale of 6440 years.

With my new list of $M_{\rm WD}$ plus the realization that a quantitative check on $\dot{M}$ is also needed, I can search for good RN candidates.  The best RN candidate is KT Eri, with a 1.25$\pm$0.03 $M_{\odot}$ WD accreting at a century-long averaged rate of 3.5$^{+1.8}_{-1.3}$$\times$10$^{-7}$ $M_{\odot}$ yr$^{-1}$ for a recurrence time scale of 40--50 years (Schaefer et al. 2021).  This result is confident enough that I claim this to be a new Galactic RN, despite having only one eruption being witnessed.  The next best RN candidate is V4643 Sgr, with a mass of 1.40 $M_{\odot}$ and a logarithmic accretion rate of -7.64 (Shara et al. 2018, see also Pagnotta \& Schaefer 2014).  The only other good RN candidates are V5589 Sgr, V1141 Sco, V1187 Sco, and V1534 Sco.

Extensive by-eye searches have been made of the Harvard and Sonneberg archival plates for old lost eruptions for the good candidates in the previous paragraph.  Even this large effort has only been able to look through many hundreds of the best plates, while missing the many lower quality plates that could easily harbor a lost eruption.  Indeed, I have discovered 14 lost eruptions on known Galactic RNe, with most of these being made on just 1--4 plates, buried deep in the archives.  Now, with the DASCH program\footnote{Digital Access to a Sky Century @ Harvard, J. Grindlay P. I., \url{http://dasch.rc.fas.harvard.edu/search.php}, see Grindlay et al. (2012).} I have exhaustively searched {\it all} Harvard plates for all 402 Galactic novae.  With this, I have found no missed eruptions from 1890 to 1989 on any of the $\sim$2500 plates for any of the 402 novae in Table 1.  This is disappointing, although it simply means that their fast  eruptions happened to have been missed.  Missed eruptions are expected to be common, as the Harvard plates coverage is typically for only 7 months each year, missing the months centered on the solar conjunction.

\subsection{Disk Versus Bulge Populations}

The sky distribution of novae consists of two overlapping distinct distributions, with one distributed closely along the entire galactic plane (the `disk' population) and the other closely distributed within a circle centered on the Galactic Center (the `bulge' population).  This dichotomy has been recognized since the 1940s (Payne-Gaposchkin 1964).  Ever since, a variety of workers have attributed a variety of demographic differences to the disk and bulge populations, always with scant evidence or discussion.  No one had even listed candidate members for each population, so tests, correlations, and demographics could not be sought or studied.  

This changed suddenly with the all-sky deep parallax survey of the {\it Gaia} astrometry satellite.  Schaefer (2018; 2022a) used the {\it Gaia} parallaxes, plus all other published evidences, to compile an exhaustive set of distances for all 402 known Galactic novae\footnote{These distances supersede all prior distance estimates, with those usually based on just one or two non-parallax measures.  Further, the {\it Gaia} parallax-only distance from the {\it Gaia} Team should not be used because they adopt priors that are inappropriate for novae in either the disk or the bulge.  This problem makes for large systematic biases for novae out past $\sim$4000 pc (Santamar\'{i}a et al. 2025).  Further, the {\it Gaia} distances do not use the large volume of additional distance measures as part of their priors, so most of the team's parallax-only measures have relatively poor accuracy.}.  With this, each nova can be attributed to be either a disk-nova or a bulge-nova.  Critical to this is the observation that the novae show a strong circular concentration with an `excess' over the disk novae that are all within 20$\degr$ of the galactic center.  The distribution has a 68\% containment radius of 5.4$\degr$.  This angle corresponds to bulge population with a Gaussian radius of 750 pc (for a Galactic Center distance of 8000 pc).  With this, 165 novae (for 41\% of all the 402 novae) are identified as being bulge-novae.  Given the uncertainties for the edge cases, a few of these 165 novae might have the incorrect population.  Suddenly, we have an exhaustive list of bulge and disk novae for use in demographic studies.

Schaefer (2022b) has systematically sought differences between the disk and bulge populations.  No significant differences were found for the nova properties of the peak absolute magnitude, the median FWHM of emission lines, the median $t_3$, the SPOC fraction, the fraction of the three spectral classes (Fe II, hybrid, and He/N), the fraction of neon novae, and the fraction of novae with subgiant companion stars.  I am impressed by the bulge and disk populations being indistinguishable (by other than their Galactic positions).  However, one parameter did have a significant difference, and that is the fraction of novae with red giant companion stars, with 5\% for disk novae and 35\% for bulge novae.  This inevitably must be due to the differing ages of the populations, with the older bulge stars having more time to evolve to red giant states.  However, detailed modeling does not reproduce the observations, at least for the first simple models.  Further, we have no explanation for why the novae in the Andromeda Galaxy with red giant companions are consistent with being entirely in the disk population (Williams et al. 2016).

Now, with my new compilation of $M_{\rm WD}$ measures, I can test whether the disk and bulge populations are different in terms of their WD masses.  For this, with 154 disk novae, the average $M_{\rm WD}$ is 1.06 M$_{\odot}$, with an RMS scatter of 0.16 M$_{\odot}$.  With 51 bulge novae, $M_{\rm WD}$ averages to 1.08 M$_{\odot}$ with an RMS scatter of 0.18 M$_{\odot}$.  That is, the bulge and disk novae show no distinction in terms of their WD masses.

\subsection{Orbital Periods}

The orbital period $P$ is the only nova property more important than $M_{\rm WD}$ for understanding the nature of each nova system in quiescence.  But for understanding the eruptions, $M_{\rm WD}$ is the one most important property.  Astronomers have spent vast efforts to measure the orbital periods of novae, dating back to first discovery that DQ Her is eclipsing back in 1959.  Schaefer (2022a) compiled a comprehensive listing of 156 reliable $P$ measures for novae.  Recently, I have used light curves from the AAVSO, {\it TESS}, {\it K2}, ZTF, and SMARTS to discover 45 new periods (Schaefer 2022a, 2025a, 2025c, 2025d), plus spectral energy distributions to estimate the first $P$ measures for 18 novae with red giant companions (Schaefer 2022a).

This list of nova periods can then be compared with my new list of $M_{\rm WD}$.  This comparison is actually a primary reason why I am writing this paper now, because I want to use it as part of a much larger analysis of all known $P$ measures for all cataclysmic variables, including dwarf nova, intermediate polars, novalike systems and more.  

For the five systems with red giant companions and periods from 227--930 days, the WD masses are over a tight range of 1.33--1.39 days.  This cannot be by chance or selection effects.  Four of the five are the well-known sister systems called `symbiotic recurrent novae' (SyRNe), including T CrB, RS Oph, V745 Sco, and V3890 Sgr.  The fifth star is V1534 Sco with its red giant companion, and which is a good RN candidate.  These five stars must arise from one evolutionary path, different from all other novae.  For some unknown reason, this evolutionary path only produces recurrent nova with WDs close to the Chandrasekhar mass.  A problem with understanding this evolutionary path is that the high accretion rate cannot be powered by the ordinary expansion of the red giant, nor can it be powered by any known mechanism for angular momentum loss.

{\bf A striking feature emerging from the $P$ versus $M_{\rm WD}$ analysis} is that amongst all systems with main sequence companion stars (i.e., $P$$<$0.7 days), only {\it one} nova has a mass $>$1.27 $M_{\odot}$.  The exception is V838 Her.  The existence of this blank area is stunning.  This blank area cannot be caused by selection effects, because any high-mass CV will have a short recurrence timescale, and hence be discovered with much higher probability than the lower-mass CVs.  And this blank area is hard to understand if the WDs are {\it increasing} in mass over each eruption cycle.  That is, if the WD mass is rising on any evolutionary time scale, then there must appear large numbers of novae of near-Chandrasekhar-mass WDs for the old systems with short period.  But these high-$M_{\rm WD}$ WDs do not exist.  Hence, the novae WDs are {\it losing} mass over each eruption cycle.  This is a strong argument that novae are not Type Ia supernova progenitors.

\subsection{Nova Shells}

\begin{figure}
	\includegraphics[width=1.01\columnwidth]{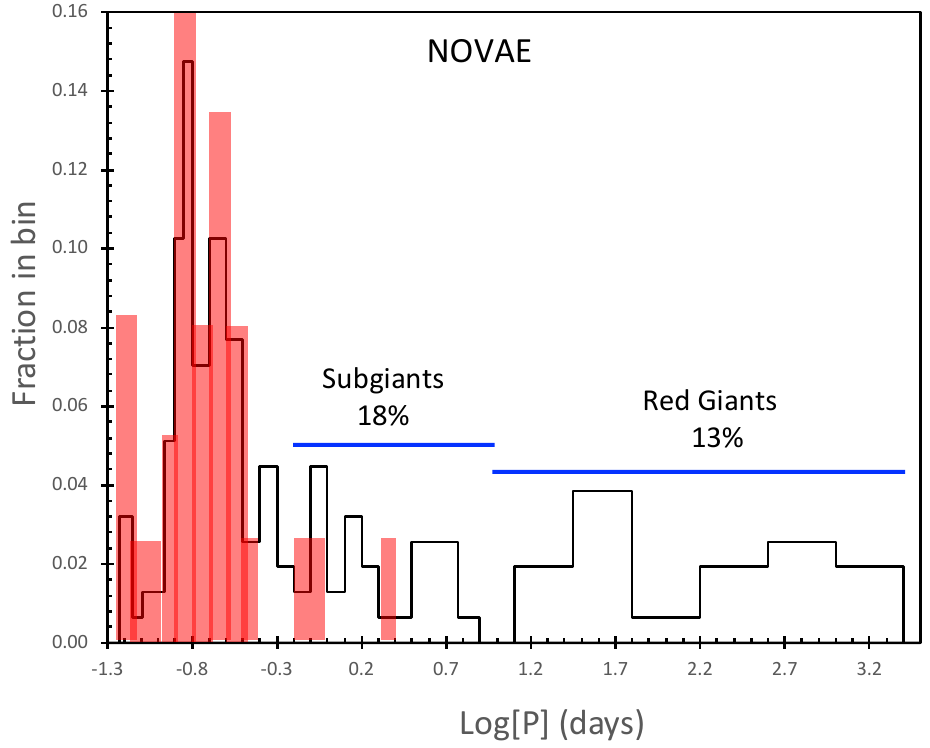}
    \caption{Histogram for shell novae and $P$.  The overall period distribution of novae is shown by the black histogram, with the base figure taken from Figure 2 of Schaefer (2022a).  The distribution of the 37 novae with both shells and $P$ measures is overplotted as a red-shaded histogram.  We see that shell-novae are strongly biased against periods longer than one-third of a day.  Normal novae have 13\% with red giant companions, whereas shell-novae have zero-percent with red giant companions.  The strong bias for bright-shells from short-orbits is apparent for novae with red giant companions, subgiant companions, and main sequence companions, so the bias does not arise from any feature of the evolutionary state of the companion (say, from a stellar wind, or surface temperature).  But I do not know any mechanism for how the shell brightness depends on the companion radius or the small semimajor axis. }
\end{figure}

\begin{figure}
	\includegraphics[width=1.01\columnwidth]{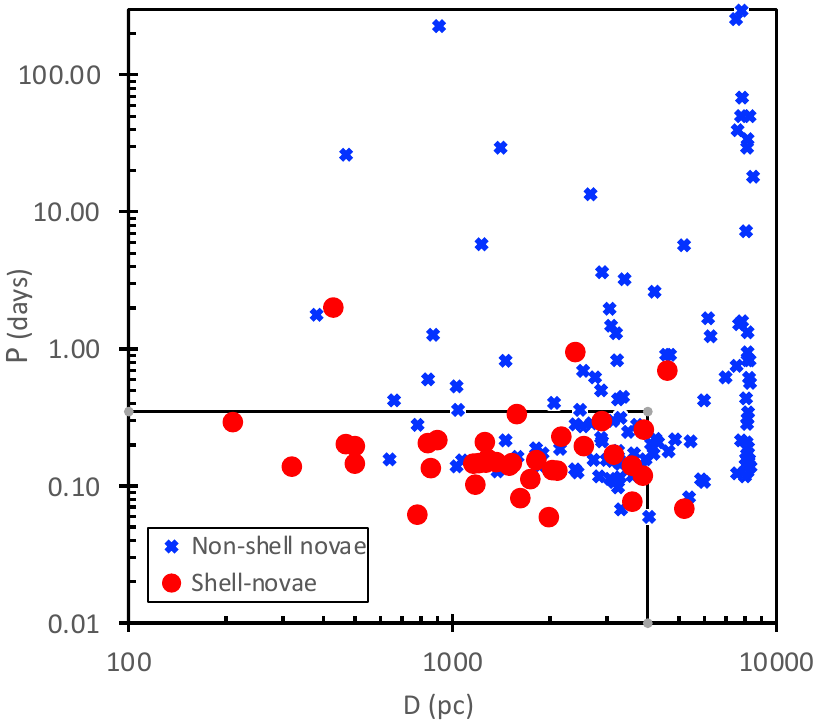}
    \caption{The distance and orbital period distribution for novae with shells and without shells.  Santamar\'{i}a et al. (2025) have already proven that distant novae rarely have visible shells, due to extinction and small unresolved sizes.  From this figure, the effective cutoff is something like 4000 pc.  Further, from Figure 6, shell-novae have a strong bias to have orbital periods shorter than one-third of a day.  These two visibility criteria make for a rectangular area in the lower left of the plot which encloses the strong cluster of shell-novae.  That is, with only four exceptions (including the famous GK Per), all the shell-novae are in the relatively small area.  This is highly significant, but I do not understand why the shells are bright for $P$$<$0.33 days. }
\end{figure}

Some novae eject shells of visible expanding gas clouds, growing larger year-after-year (Downes and Duerbeck 2000, Santamar\'{i}a et al. 2025).  This is just the nova ejecta, separated enough from the inner binary so as to be resolved and bright enough to be visible in its own right.  The first discovered nova shell\footnote{This is not to be confused with the light echo of the eruption light reflecting off dust grains in directly intervening clouds, with this being visible even in the tail of the eruption.  The first discovered light echo was in front of GK Per.  Further, this is not to be confused with the claimed shell around T Aur reported by E. E. Barnard in 1892, as this turned out to just be caused by unfocused blue light.} was around GK Per (Shara et al. 2012), with this still being brightly visible by eye even today with amateur scopes.  My big list identifies the 49 known nova shells.  This leaves a large number of other novae for which no shell has been recognized, so even though the expanding shells must be surrounding the central star, for some reason they are not bright enough to be detected.  A primary question is why some novae display visible shells while others do not?  This is a demographics question, with one of the best ways to address this is by seeking correlations between the nova properties in Table 1.

Another primary question for nova shells is to learn the physics for what makes for a bright shell?  I am not aware of any paper reporting any theory or model that calculates or speculates on the physics of nova shell brightnesses.  With this absence, the only approach can be observational and empirical, and must involve comparisons of novae with visible shells with novae with no detected shells.  So again, we have a primary question for nova demographics.  The wonderful paper of Santamar\'{i}a et al. (2025) constructed a full catalog of known nova shells and sought correlations with many nova properties.  They found no significant distinction between nova with and without shells as based on the properties of $t_3$, age of the shell, expansion velocity of the shell, and the Galactic coordinates.  They also find a strong correlation with $D$, saying ``Compared to the general population of Galactic novae, resolved nova remnants are found preferentially at closer distances, as the lower extinction and larger H$\alpha$ surface brightness favour their detection.''

I can extend the nice work in Santamar\'{i}a et al. (2025) by seeking correlations with other nova properties included in Table 1.  One of these newly-tested properties is the WD mass.  I could imagine that the shell brightness is correlated with $M_{\rm WD}$.  The WD mass is strongly correlated with the ejecta mass and velocity (c.f. Yaron et al. 2005), such that a low-mass WD will have a massive ejecta expanding slowly, making for a high-density and a presumably high-brightness.  This is what makes for the big dust dips in the D-class light curves.  Nevertheless, the nova shell systems have WD masses ranging from 0.55 M$_{\odot}$ (for V723 Cas) to 1.27 M$_{\odot}$ (for T Pyx).  This distribution of WD masses is indistinguishable from the general nova population.

I can expect that shells will be brighter for more massive ejecta and slow ejecta that makes for higher density in the shell.  With this, the JDF light curve classes should have a higher fraction for visible shells.  In particular, the same dense ejecta that creates the dust clouds of D-class novae should also make for brighter shells.  If this expectation is correct, then the novae with shells should have a large fraction of JDF novae (as compared to the fraction for non-shell novae).  This can be tested.  Out of the 41 novae with shells and light curve classifications, 12 are of the D-class (29\%), while Strope, Schaefer, \& Henden (2010) give the fraction to be 18\% for bright novae, with this difference not being of any high significance.  The massive and slow ejecta of JDF novae are 22 out of 41 (54\%) of the novae with visible shells, as compared to 36\% for the Strope sample, with this difference not being adequate to declare a significant effect.  The effect of expansion velocity can be tested directly by comparing the average FWHM for novae with shells, 2000 km/s, versus the average FWHM for novae without shells, 1630 km/s.  The difference is not of any useable significance, and it actually works against my simple expectation.  In all, I have found no significant evidence that shell brightness depends on expansion velocity or light curve class.

I also expect that the shell brightness might be correlated with spectral properties.  For two examples, maybe the higher excitation state of the ejecta in a He/N nova will make the late-time ionization state conducive to brighter emission lines, or maybe the ejecta from a neon nova will have a relatively small hydrogen composition and a faint H$\alpha$ line.  For the 41 novae with a shell and a spectral class assignment, 5 have a He/N classification (12\%), while 18\% of the non-shell novae are He/N, with this difference not being of any high significance.  For the 42 novae with a shell and a spectral information, 14 have been identified as neon novae (33\%), while 14\% of the non-shell novae are neon novae, with this difference not being of any high significance.   So I am not seeing any actionable correlation between shells and spectral properties.

I had no expectations that the shell brightness would be effected by the orbital period.  But a casual glance at my listing of the 38 shells with measured $P$ shows that none of them have red giant companions.  This is in contrast to the overall frequency of 13\% (Schaefer 2022a).  The Poisson probability of zero red giant companions out of 38 shells is $(1-0.87)^{38}$ = 0.44\%, getting close to a 3-sigma confidence level.  For subgiant companion stars (i.e., with $P$ from 0.6--10 days), there are 4 shell novae, including the famous GK Per at 1.9968 days and the exceptional-in-everything V445 Pup at 1.8735 days.  

Currently, the largest and brightest nova shell is from the unique V445 Pup with a subgiant companion star.  V445 Pup is in a class by itself, being the only known helium nova, for which the eruption properties and evolution are greatly different from those of all the other hydrogen novae in this study (Kato et al. 1989; Kato \& Hachisu 2003).  V445 Pup is notorious for having a bright bilobate fast-expanding (6720 km s$^{-1}$) nova shell and for having an incredibly deep and long-lasting dust dip.  V445 Pup is also one of the highest mass WDs known in a nova, at $>$1.35 M$_{\odot}$ (Kato \& Hachisu 2003).  In its 2000 eruption, the V445 Pup WD ejected $\gg$0.001 M$_{\odot}$, which is much more than accreted during the previous inter-eruption interval, so the WD is losing mass over each eruption cycle (Schaefer 2025a).  V445 Pup has five extreme properties; zero hydrogen, by far the deepest and longest dust dip, the largest ejecta mass, one of the fastest ejecta velocities, and one of the highest WD masses.  Although with only a sample of one, it seems that these five coincident extreme properties should be causally connected, with the hydrogen-deficiency taking primacy for producing the other extreme properties.

The shell-novae have no red giants, few subgiants, and few main sequence systems with $P$$>$0.33 days.  It looks like there is a strong preference for shell-novae to have short periods.  This can be quantified by a Kolmogorov-Smirnov test asking whether the period distributions for shell-novae and non-shell-novae come from the same parent population.  By this test, the shell-novae are different from non-shell-novae at the 99.89\% confidence level.  Figure 6 shows a histogram of shell-nova periods, superposed on a histogram of non-shell-nova periods.  The bias against $P$$>$0.33 day nova with bright shells is easy to see in this plot.

The two strong biases on $D$ and $P$ work together, as can be seen in Figure 7.  Of the 37 shell-novae with measured $P$, 33 are inside a region bounded by $P$$<$0.33 days and $D$$<$4000 pc.  This can be compared to the non-shell-novae, with only 34\% (out of 131) inside the region.  If the shell-novae follow the distribution of the non-shell novae, the probability of having four or fewer outliers amongst the shell-novae is 4$\times$10$^{-7}$.  So the cluster and the bias seen in Figure 7 is highly significant.

The outliers in Figure 7 are GK Per ($P$=1.99680 d), CP Cru (0.944 d), V723 Cas (0.69327 d), and V458 Vul (0.06812 d).  V458 Vul is the only nova that has recently emitted a classic planetary nebula, V723 Cas is a V1500 Cyg star and is even now far brighter than its pre-eruption level, and GK Per is an intermediate polar.  But I am seeing no consistent property that can account for them being outliers.

Figure 7 also shows many non-shell-novae deep inside the rectangular area, so we have to wonder why they do not show visible shells?  The five most extreme cases are V1369 Cen ($P$=0.15656 days), V728 Sco (0.13834 d), CP Lac (0.145143 d), V1500 Cyg (0.139617 d), and IL Nor (0.06709 d).  V1500 Cyg is the only nova that is an asynchronous polar, and it has a fantastic pre-eruption rise for a month before the extremely fast eruption.  V1369 Cen is a $\gamma$-ray nova.  IL Nor is one of the few novae with a period {\it below} the nova period gap.  However, there is no nova property that is consistent throughout the non-shell-novae inside the shell-region.  So I have no explanation for why these novae do {\it not} show any visible shell.

The strong bias for novae with detected shells to be closer than 4000 pc follows expectations for brighter and more visible shells.  But detectability depends on the surface brightness, not on the overall brightness, and the surface brightness is distance independent.  Santamar\'{i}a et al. (2025) correctly solve this issue by invoking the extinction increasing with distance.  Further, the shell angular radius decreases with distance, making for the shells to be unresolved until later times, after which they have faded.

The novae with detected shells have a strong bias to have orbital periods $<$0.33 days.  I have no explanation for this bias.  The mechanism creating this bias cannot arise from any property associated with the evolutionary state of the companion star (e.g., its surface temperature, stellar wind strength) because the same bias exists for all of the novae with red giant companions, with subgiant companions, and with main sequence companions.  I can well imagine that the bias mechanism could be determined by the semimajor axis of the companion star, with a small-$P$ allowing the companion to be deep inside the quasi-stationary envelope around the WD, whipping up the mass ejection to form a bright shell.  But such speculation needs a real physical model calculation.

\subsection{Gamma-Ray Novae}

An exciting and unexpected discovery of the last fifteen years has been that some novae are emitting GeV $\gamma$-rays (Chomiuk, Metzger, \& Shen 2020).  These were all detected with the $\it Fermi$-LAT detector seeing flux from 0.1--10 GeV.  Indeed, V959 Mon was first discovered with $\gamma$-rays, around the time of solar conjunction.  The mechanism for gamma-ray production is the collisions between hadrons, like the protons within the ejecta hitting other protons within the ejecta, producing a $\pi^0$ that quickly decays into two high-energy photons.

The first discovered $\gamma$-ray nova was V407 Cyg, for which the Mira companion star suggests that the gamma radiation is produced by the nova ejecta shocking into the dense wind of the Mira star.  However most subsequent detections were for novae with main sequence companions that cannot have any dense wind or circumstellar shell.  This is taken to mean that the dominant radiating shocks are {\it internal}, where varying layers within the ejecta have differing velocities and so collide with each other.  (This is separate from {\it external} shocks, where the ejecta hits surrounding low velocity gas in the vicinity.)  The existence and dominance of internal shock is strikingly demonstrated by the three jitter flares around the peak of V906 Car being simultaneous with three $\gamma$-ray flares\footnote{These three optical/$\gamma$ flares are also a strong clue that the jitters are caused by internal shocks, with this being the long-needed break in starting to understand the jitter phenomenon.} (Aydi et al. 2022).

Outstanding questions include whether the red giant systems (`RedG') are different in some way (like higher $\gamma$-ray luminosity, $L_{\gamma}$) from the novae with main sequence companions (`MainSeq') due to their extra external shocks, whether the roughly-10\% of $\gamma$-novae are somehow a separate population, and whether the various nova properties are correlated with $L_{\gamma}$?  These are demographic questions, which benefit from large numbers of nova detections and non-detections, as well as having a full catalog of nova properties.  Chomiuk et al. (2020) conclude that the {\it Fermi} detections are correlated with distance with only `marginal significance', and further that ``no clear correlations have been found between gamma-ray luminosity and nova properties".  In this paper, I have doubled the number of $\gamma$-novae, I have greatly improved distances, and I have a full catalog of the nova properties. 

\subsubsection{Demographics of nova shocks}

\begin{table*}
	\tablenum{4}
	\centering
	\caption{$\gamma$-ray nova properties}
	\begin{tabular}{llrllrlrrlrr}
		\hline
		Nova 	&	LC class  & $V_{\rm peak}$ & Spec class & Neon  &  FWHM   &  $P$  &  $M_{\rm WD}$  &  $D$  &  Companion   &   $F_{\gamma}$ &  $\log[L_{\gamma}$] \\
		\hline
V1369 Cen	&	D(65)	&	3.3	&	Fe II	&	...	&	408	&	0.15656	&	0.91	&	640	&	MainSeq	&	2.1	&	37.01	\\
V572 Vel	&	...	&	4.8	&	...	&	...	&	...	&	0.12318	&	...	&	1276	&	MainSeq	&	6.4	&	38.10	\\
V5668 Sgr	&	D(78)	&	4.3	&	Fe II	&	...	&	2800	&	0.15616	&	0.78	&	1290	&	MainSeq	&	0.75	&	37.17	\\
V339 Del	&	PP(29)	&	4.8	&	Fe II	&	...	&	1421	&	0.16294	&	1.18	&	1590	&	MainSeq	&	2.1	&	37.80	\\
V1405 Cas	&	J(175)	&	5.2	&	Hybrid	&	Neon	&	...	&	0.18839	&	0.53	&	1810	&	MainSeq	&	1.4	&	37.74	\\
V549 Vel	&	J(118)	&	9.1	&	Fe II	&	...	&	...	&	0.40317	&	0.64	&	2060	&	MainSeq	&	0.8	&	37.61	\\
YZ Ret	&	P(22)	&	3.7	&	Hybrid	&	Neon	&	2500	&	0.13245	&	1.24	&	2390	&	MainSeq	&	6.0	&	38.61	\\
V462 Lup	&	S	&	5.4	&	...	&	...	&	...	&	0.07489	&	...	&	2666	&	MainSeq	&	3.2	&	38.43	\\
RS Oph	&	P(14)	&	4.8	&	He/N	&	Neon	&	3930	&	453.6	&	1.33	&	2710	&	RedG	&	24	&	39.32	\\
V959 Mon	&	S	&	$<$9.4	&	...	&	Neon	&	...	&	0.29585	&	1.05	&	2900	&	MainSeq	&	5.2	&	38.72	\\
V407 Lup	&	S(8)	&	6.4	&	Fe II	&	Neon	&	3700	&	3.62	&	1.20	&	2900	&	MainSeq	&	1.8	&	38.26	\\
V1716 Sco	&	S(12)	&	7.3	&	Fe II	&	...	&	2600	&	1.35101	&	1.21	&	2914	&	MainSeq	&	6.5	&	38.82	\\
V357 Mus	&	D(32)	&	6.7	&	...	&	...	&	...	&	0.15516	&	0.94	&	2990	&	MainSeq	&	2.4	&	38.41	\\
V407 Cyg	&	P(44)	&	7.9	&	He/N	&	...	&	2670	&	...	&	1.22	&	3100	&	RedG	&	6.6	&	38.88	\\
V5856 Sgr	&	P(14)	&	5.9	&	Fe II	&	...	&	...	&	...	&	...	&	3180	&	MainSeq	&	9.7	&	39.07	\\
V1674 Her	&	S(2)	&	6.2	&	Hybrid	&	Neon	&	5600	&	0.15302	&	...	&	3220	&	MainSeq	&	55	&	39.83	\\
FM Cir	&	J(85)	&	5.9	&	Fe II	&	...	&	1500	&	0.14977	&	0.66	&	3250	&	MainSeq	&	1.3	&	38.22	\\
V392 Per	&	P(11)	&	6.2	&	...	&	Neon	&	...	&	3.21997	&	...	&	3400	&	MainSeq	&	8.9	&	39.09	\\
V906 Car	&	J(64)	&	5.9	&	Hybrid	&	...	&	325	&	...	&	...	&	3720	&	MainSeq	&	36	&	39.78	\\
V679 Car	&	S(24)	&	7.8	&	Fe II	&	...	&	2000	&	0.61975	&	0.98	&	6970	&	MainSeq	&	1.9	&	39.04	\\
V6598 Sgr	&	C(9)	&	10.2	&	...	&	...	&	4000	&	...	&	...	&	7661	&	MainSeq	&	1.9	&	39.13	\\
V1535 Sco	&	P(20)	&	9.5	&	He/N	&	...	&	2000	&	50	&	0.85	&	7790	&	RedG	&	1.0	&	38.86	\\
V1324 Sco	&	D(30)	&	10.0	&	Fe II	&	...	&	2000	&	...	&	0.80	&	7870	&	MainSeq	&	5.0	&	39.57	\\
V745 Sco	&	P(9)	&	9.4	&	He/N 	&	...	&	3600	&	930	&	1.39	&	8020	&	RedG	&	3.0	&	39.36	\\
V1707 Sco	&	S(7)	&	11.8	&	Fe II	&	...	&	6900	&	...	&	...	&	8040	&	MainSeq	&	2.3	&	39.25	\\
V5855 Sgr	&	J(19)	&	7.8	&	Fe II	&	...	&	...	&	...	&	...	&	8050	&	MainSeq	&	2.6	&	39.30	\\
V1723 Sco	&	S(14)	&	7.0	&	Fe II	&	...	&	...	&	...	&	1.12	&	8080	&	MainSeq	&	19	&	40.17	\\
V3890 Sgr	&	S(14)	&	8.1	&	Hybrid 	&	...	&	2140	&	747.6	&	1.37	&	8550	&	RedG	&	1.2	&	39.02	\\
		\hline
28 $\gamma$	&	63\%	&		&	18\%	&	25\%	&		&		&		&		&	18\%	&		&		\\
~~~novae:	&	SPOC	&	6.6	&	He/N	&	neon	&	2550	&	0.242	&	1.05	&	3140	&	RedG	&	2.8	&	38.87	\\
		\hline
	\end{tabular}
\end{table*}

An estimated $\sim$10\% of discovered events are $\gamma$-novae.  I have collected the properties of all the 28 currently-known $\gamma$-novae (Franckowiak et al. 2018, Chomiuk et al. 2020, Wang et al. 2024) into Table 4.  These are sorted with increasing distance $D$.  In addition, there are six novae\footnote{Table 1 was constructed from the base table in Schaefer (2022b) that was complete only for novae peaking before mid-2021.  But five of the gamma-detected novae are after mid-2021, so I should include them as a substantial fraction of the known population.  With this, I have added V462 Lup, V6598 Sgr, V1716 Sco, V1723 Sco, and V572 Vel to Table 4.  Further, V407 Cyg was mistakenly not included in Schaefer (2022b), so I now include it in Table 4.} not in Table 1, for which I have correctly calculated the best distances and have added various properties.  The first ten columns are extracts from Table 1.  The second from last column gives the $\gamma$-ray flux, $F_{\gamma}$, in units of $10^{-7}$ photons/cm$^2$/s for photon energies $>$100 MeV.  The last column gives the logarithm of the gamma-ray luminosity, $L_{\gamma}$, in units of photons per second, calculated as $4\pi D^2 F_{\gamma}$.  The bottom two-line section of Table 4 gives a summary for each column, variously the number of novae, the percentage for some class in the column, or the median of the values in the column.

I have 5 $\gamma$-novae with red giant companion stars (RS Oph, V3890 Sgr, V1535 Sco, V745 Sco, and V407 Cyg).  V407 Cyg is the first discovered $\gamma$-novae, and it has a Mira star companion with a very high stellar wind.  RS Oph, V745 Sco, and V3890 Sgr are three of the four famous sister RNe\footnote{The fourth sister is the most famous of all novae, T CrB, which will presumably have its fifth observed eruption any month now.  With a distance of 910 pc and $t_3$=6 days, I predict that T CrB will rise up to a $\gamma$-ray flux of 350 times $10^{-7}$ photons/cm$^2$/s.  This is two orders-of-magnitude brighter than almost all {\it Fermi} nova.}.  V1535 Sco is an ordinary nova in our Galactic bulge, for which four indicators point to the WD mass being $\sim$1.16 $M_{\odot}$, rather than the tabulated 0.85 $M_{\odot}$, see footnote 7.  If external shocks are significant, these are the five systems where they should dominate.  These red giant systems have an average $\log L_{\gamma}$ of 46.09 with an RMS scatter of 0.24.  The main sequence systems have an average of 45.66 with an RMS scatter of 0.84.  A Kolmogorov-Smirnov test returns a probability of 0.15 that the two observed $L_{\gamma}$ distributions are from the same parent population.  (With a two-sided Gaussian, this probability corresponds to what could be called a 1-sigma difference.)  So formally, the red giant systems and the main sequence systems appear to have the same luminosity function.

Importantly, V407 Cyg has a Mira star and is a D-type symbiotic star with a very heavy stellar wind and dust shell, while the other four red giant systems are S-type symbiotic stars with greatly weaker stellar winds.  The $L_{\gamma}$ should be proportional to the density of the wind for any external shocks.  But V407 Cyg is at the bottom of the luminosity function for red giant systems.  This means that the high density of the V407 Cyg wind cannot be contributing much luminosity, and that the contribution from external shocks is negligibly small.  This is also seen with $L_{\gamma}$ for V407 Cyg being right in the middle of the luminosity function for the 23 main sequence systems.  That is, V407 Cyg has no significant excess flux due to its external shocks.  So from the last two paragraphs, we have a confident answer that external shocks (with the nova ejecta impacting the red giant stellar wind) are contributing only a negligible $\gamma$-ray flux.

For the internal shocks (with shells within the ejecta colliding with each other), we have the proof from the blatant simultaneity of the three $\gamma$-ray flares and the three optical jitters in V906 Car (Aydi et al. 2020).  So far, only 5 $\gamma$-novae of the J-class (with jitters) have been discovered, V906 Car, V1405 Cas, FM Cir, V5855 Sgr, and V549 Vel.  Other than the great result for V906 Car, I can check the other 4 J-class $\gamma$-novae for correlations between optical jitters and $\gamma$-ray flares:  For FM Cir, the optical light curve shows three 1--2 mag jitters at times when the {\it Fermi} light curve looks to be flat (Wang et al. 2024).  For V1405 Cas, Buson et al. (2021) give an inadequate verbal description of the $\gamma$-ray light curve, reporting many null detections up until 2021 May 20, but the one huge jitter peaked 11 days earlier and the optical light curve shows the jitter to have been over for 2 days at the time of the {\it Fermi} discovery, so optical and $\gamma$ light curves appear to be anti-correlated.  For V5855 Sgr, Li \& Chomiuk (2016) report many null detections over the first big jitter, only to have detected $\gamma$-ray flux only for the first three days of a 7-day jitter, missing its peak.  For V549 Vel, {\it Fermi} detected flux only on days that were far from any of the four big optical jitters, while the big optical jitter were invisible in the $\gamma$-rays, for a strong {\it anti-correlation} between optical/$\gamma$ flux.  Li et al. (2020) conclude ``The optical and $\gamma$-ray light curves of V549 Vel show no correlation'' and ``the optical and $\gamma$-ray light curves of V549 Vel show no correlation, likely implying relatively weak shocks in the eruption.''  That is, for all J-class novae (other than V906 Car), the jitters do not correspond in any way to any $\gamma$-ray flares.  So the nice and convincing V906 Car result is anti-confirmed four times over.  I can only think that the other four jitter-novae have some unrecognized circumstance confusing the issue.  But even that is just saying that most internal shocks in novae do not provide any substantial fraction of $L_{\gamma}$.

I can test whether $L_{\gamma}$ for J-class novae have a different parent population than other $\gamma$-novae.  With the Kolmogorov-Smirnov test, the returned probability is 0.45, which is around that expected for the two populations being identical.  That is, J-class novae are not significantly more luminous than all the other novae.  So the isolated optical jitters are not producing any apparent extra $L_{\gamma}$.  

If we now try to claim that internal shocks are the dominant producers of $L_{\gamma}$, then we have to wonder why all the other novae produce the same $\gamma$-flux {\it without} showing jitters?  That is, $\gamma$-novae of the S, P, O, C, and F classes have finely smooth light curves, with a smoothness that is impossible from the shot-noise of many days-long jitters.  It is not satisfying to presume that all non-J-class novae have their internal shocks made frequently or regularly such that the optical light curve is smooth.  An alternative answer is that the jitter-shocks do not produce much gamma-flux.

Presumably the $L_{\gamma}$ is proportional to the kinetic energy in the ejecta, so $L_{\gamma}$$\propto$$M_{\rm ejecta}$$FHWM^2$.  Low-mass WDs eject large masses at low velocities, while high-mass WDs ejecta small masses at high velocities.  So it is unclear which effect dominates.  I can seek an empirical answer by comparing $L_{\gamma}$ for the JDF and SPOC $\gamma$-novae.   For this, $\langle \log[L_{\gamma}] \rangle$ is 45.98 for SPOC and 45.31 for JDF, suggesting that the SPOC novae are around 4.7$\times$ the gamma-luminosity as JDF novae.  However, the Kolmogorov-Smirnov test returns a probability of 0.19, which is to say that the SPOC and JDF luminosity functions are not significantly different.

I can directly test whether $L_{\gamma}$ is correlated with FWHM.  This is testing the presumption that the differential velocities between shells of the ejecta are proportional to the shell velocity measured by the FWHM.  For this, the correlation statistic for $L_{\gamma}$ versus the logarithm of the FWHM shows no apparent correlation or trend.  We expect factor-of-2 measurement errors for both $L_{\gamma}$ and FWHM, but the scatter is over one order-of-magnitude.  So the scaling as FWHM$^2$ is not seen in my demographic data.

\begin{figure}
	\includegraphics[width=1.01\columnwidth]{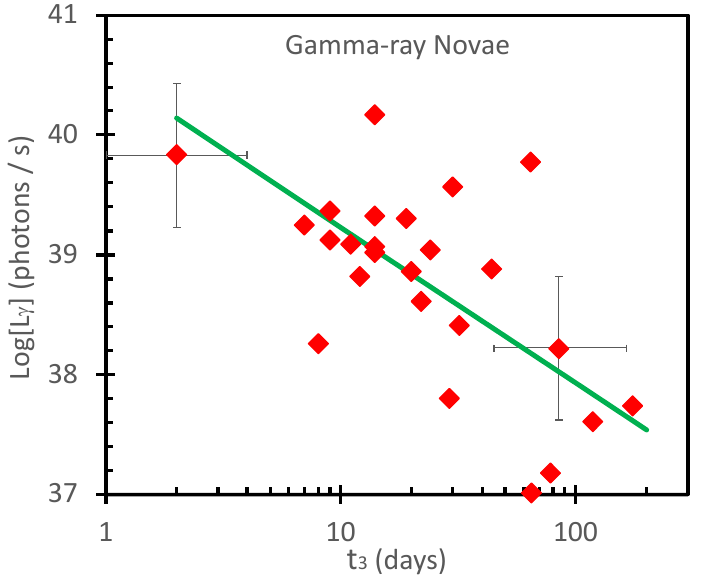}
    \caption{Gamma-ray luminosity versus $t_3$.  For the 25 $\gamma$-ray novae with measured decline rate $t_3$ (see Table 4), the gamma-ray luminosity is correlated with the $t_3$.  Typical measurement error bars are shown for two novae.  The scatter about the best fit power law (the green straight line) is large.  The relation is significant at the 4.2-sigma confidence level.  This correlation is quantified with equation 5.  So, for some unknown reason, the $\gamma$-bright novae are the fast novae, with relatively high $M_{\rm WD}$. }
\end{figure}

A secondary indicator of the ejecta density and velocity is the decline rate, $t_3$, so I should seek a correlation with $L_{\gamma}$.  Unexpectedly, I have found a poor-but-significant correlation, with the slowest $\gamma$-novae being nearly two orders-of-magnitude less-luminous than the fastest novae.  A log-log plot of $L_{\gamma}$ versus $t_3$ (see Figure 8) shows a clear trend, although the scatter is large.  (We expect typical measurement errors of factor-of-two for both $L_{\gamma}$ and $t_3$, but the scatter in the plot is larger.)   A chi-square fit to the 25 $\gamma$-novae with $t_3$ measures in Table 4, with an adopted uncertainty in $\log L_{\gamma}$ of 0.6, returns a power law index of -1.3$\pm$0.3.  So the best fitting correlation returns 
\begin{equation}
\log [L_{\gamma}]=40.53-1.3\log[t_3], 
\end{equation}
with units of photons/sec.  This correlation is significant and real, so should provide a good clue to the physics of the high-energy luminosity.  But I have no real insight for why the nova duration matters while the FWHM does not.  More in particular, I do not understand why a slow $\gamma$-nova should be greatly fainter in $L_{\gamma}$.

I can seek correlations between $L_{\gamma}$ and other intrinsic nova properties.  The $M_{\rm WD}$ distribution of the $\gamma$-novae (stretching from 0.53 to 1.39 $M_{\odot}$) is similar to the distribution for all novae, while there is no correlation between $L_{\gamma}$ and $M_{\rm WD}$.  The orbital period distribution of the $\gamma$-novae (stretching from 0.07489 to 930 days) is indistinguishable from that of all known nova periods (Schaefer 2022a), and has no significant correlation with $L_{\gamma}$.  The $\gamma$-nova have 25\% (7 out of 28) recognized as being a neon nova, with this fraction similar to the fraction for all novae, so there is no significant effect on $L_{\gamma}$ of whether the WD is of CO or ONe composition.  The $\gamma$-nova have 18\% (4 out of 22 measured) with spectral class He/N, with this fraction identical to that for all disk novae (Schaefer 2022b), so the $\gamma$-ray luminosity is independent of the spectral class.

\subsubsection{Visibility of $\gamma$-novae}

\begin{figure}
	\includegraphics[width=1.01\columnwidth]{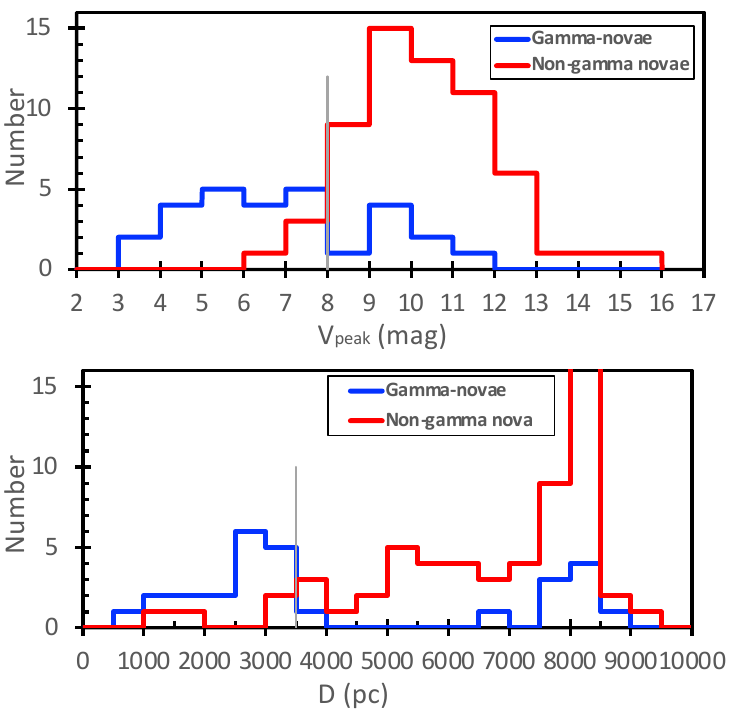}
    \caption{Histograms for the $V_{\rm peak}$ and $D$ distributions.  The two panels show the distributions for the 28 $\gamma$-nova (in blue) and the 95 non-$\gamma$ novae from 2010.0 to 2022.6.  The top panel shows the distribution of the nova peak magnitude.  We see that both distributions have sharp changes at $V$=8.0.  Almost all novae brighter than 8th mag are detected by {\it Fermi}, while most nova fainter than 8th mag are not detected.  That is, optically bright novae are usually detected by {\it Fermi}.  The bottom panel shows the distribution of the nova distances, $D$.  We see a threshold at 3500 pc, where nearer novae are usually discovered by {\it Fermi} and farther novae are usually not detected.  The primary exception for this is that a modest fraction of the novae in the Galactic bulge (with $D$$\sim$8000 pc) are detected.  But all these bulge detections are for the fastest novae (with $t_3$$\le$30 days), for which the $L_{\gamma}$ are the brightest.  This demonstrates that the visibility of novae by {\it Fermi} is controlled by {\it both} $D$ and $t_3$ (see Figure 10 and Equation 6). }
\end{figure}

\begin{figure}
	\includegraphics[width=1.01\columnwidth]{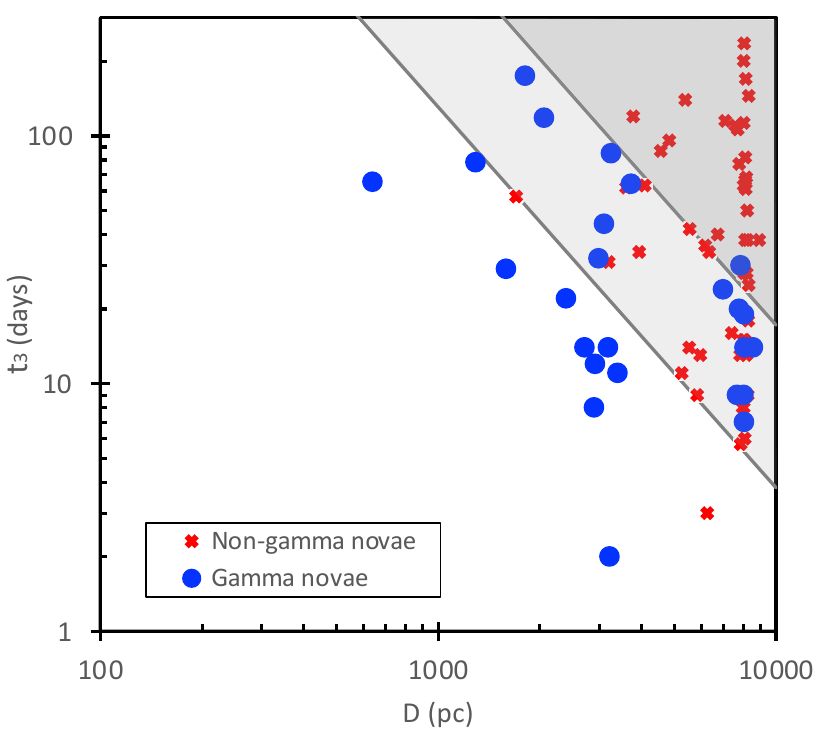}
    \caption{Detectability of $\gamma$-ray novae by {\it Fermi}.  In this plot of $t_3$ versus $D$, the 28 $\gamma$-novae are represented by the large blue circles, while the non-detections (from 2010.0 to 2022.6) are represented by the small red $\times$ symbols.  The visibility of novae depends on the gamma-ray flux, which is $L_{\gamma}/(4\pi D^2)$, for which $t_3$ is a proxy for the luminosity through Equation 5.  The {\it Fermi} threshold for $>$100 MeV flux from a nova is observed to vary between 0.8 and 5.0 (in units of $10^{-7}$ photons cm$^{-2}$ s$^{-1}$).  These limits are represented by the two parallel lines of constant flux, $F_{\gamma}$, connecting from the middle of the top down to the middle of the right side.  The unshaded-region below the bottom line is for novae that are above the highest of the detection thresholds, and the darkly-shaded region to the upper-right of the upper line is for bursts that are fainter than the lowest thresholds.  The middle lightly-shaded region is for the variable detection threshold of {\it Fermi}, where some novae will be detected and some novae will be missed.  This figure presents the explanation for why {\it Fermi} only discovers 20\% of novae.  Further, this figure demonstrates that $\gamma$-novae are not a separate class of novae, rather, they are just the novae that are near enough to be detected.}
\end{figure}

The only outstanding question from above is why only $\sim$10\% of novae\footnote{Actually, my tabulation for the years 2010.0 to 2022.6, has cataloged 118 novae, of which 23 are $\gamma$-novae, for a fraction of 20\%.} are detected by {\it Fermi}?  This is partly asking whether the $\gamma$-novae constitute a distinct subset out of all novae, or are simply the brightest by chance?  This is a demographics question, for which prior reviews (e.g., Franckowiak et al. 2018, Chomiuk et al. 2020) have come to no useable conclusion.  Now, with double the number of $\gamma$-novae, greatly better distances, and a full set of nova properties, I can answer the questions of visibility.

The detection of $>$100 MeV $\gamma$ radiation from novae is patchy, with no clear explanation for why some novae are detected and some are not.  The obvious factor is the peak magnitude of the nova, with $\gamma$-novae often being relatively bright.  This would be using the $V_{\rm peak}$ as a proxy for the nova distance $D$, with this critically entering the question of the gamma-ray flux through the usual inverse-square law of photon brightness.  Novae in general are moderately good `standard candles', with their peak absolute magnitudes, $M_{\rm V,peak}$, always being near $-7.45\pm1.33$ (Schaefer 2022b).  The top panel of Figure 9 shows a histogram of $V_{\rm peak}$ for all 28 $\gamma$-nova.  Overplotted is the histogram for 95 non-$\gamma$ novae peaking from 2010.0 to 2022.6.  This histogram shows a sharp demarcation at $V$=8.0, with most of the brighter novae being detected by {\it Fermi}, and with most of the fainter novae {\it not} being detected by {\it Fermi}.  So to first order, a nova peak brighter than 8th mag will likely be detected, and novae peaking fainter than 8th mag will likely not be detected.

This plot has fairly poor predictive power.  Part of the trouble is that the plotted $V_{\rm peak}$ measures have not been corrected for extinction.  Part of the problem is that the RMS scatter of $M_{\rm V,peak}$ is 1.33 mag, so we are including all the variations of the optical luminosity function into the distance.  Rather, with my full set of distances for all 402 Galactic novae, we should just use the $D$ measures rather than a proxy.  Part of the problem is that 4 non-$\gamma$ novae violate the threshold by a little, and 8 $\gamma$-novae violate the threshold by a lot.

The $\gamma$-nova violators are all bulge novae, nearly at $D$$\approx$8000 pc.  These distinguish themselves from other bulge novae in that they have fast light curves, with $t_3$$\le$30 days.  From Equation 5, we know that these novae in particular have high $L_{\gamma}$ values.  So we have a schematic explanation for the bulge violators, they are near the top of the gamma-ray luminosity function, so they are visible even in the bulge.  The rest of the bulge novae have longer $t_3$, lower luminosity, and thus were never detected.

A better visibility criterion is to go directly to my $D$ measures.  The bottom panel of Figure 9 shows the histogram for distances.  Again, as expected, there is a sharp demarcation, which is seen to be at 3500 pc.  That is, most novae nearer than 3500 pc are detected by {\it Fermi}, while most novae farther than 3500 pc are {\it not} detected by {\it Fermi}.

But the distance histogram still has $\gamma$-nova outliers at the distance of the bulge.  Again, the explanation is that these are the fastest novae, so by Equation 5, they are the most luminous, and they can be detected even in the bulge.  As a guide for future novae, we can expect a {\it Fermi} detection if the nova is $D$$<$3500 pc, or if it has $t_3$$<$30 days.

We can do better, by having a joint criterion using both $D$ and $t_3$.  For this, Figure 10 is a plot of all the {\it Fermi} detection on a $t_3$ versus $D$ plane.  Now we see that the {\it Fermi} detections are all to the lower left in the plot and the {\it Fermi} non-detections are all to the upper right in the plot.  There is a mixing region as a middle slanted stripe where the detections and non-detections are mixed up.  The width of this middle-stripe arises because the effective {\it Fermi} detection threshold varies substantially.

We can quantify the observed flux, $F_{\gamma}$ in units of $10^{-7}$ photons cm$^{-2}$ s$^{-1}$, by combining $L_{\gamma}$=$4\pi D^2 F_{\gamma}$ with Equation 5, to get
\begin{equation}
F_{\gamma} =(10^{40.53} t_3^{-1.3})/(4\pi D^2). 
\end{equation}
Curves of constant flux appear as straight lines in the log-log plot of Figure 10, with two examples plotted parallel from upper left to lower right.

The detection threshold for {\it Fermi} varies by nearly an order-of-magnitude.  Franckowiak et al. (2018) give lists of detection flux limits for many undiscovered novae and detection fluxes for many $\gamma$-novae.  Table A.1 of their paper shows that $\gamma$-novae can be detected down to 0.75, 0.8, and 1.0 in units of $10^{-7}$ photons cm$^{-2}$ s$^{-1}$.  Their Figures 17 and 19, and Table 3 show that their flux limits for nova discovery can get up to 3--7 in the same units.  So the flux range of uncertain detection for {\it Fermi} is close to 0.8 to 5.0 in the same units.  The two lines drawn in Figure 10 correspond to these flux limits.

We now have an empirical explanation for which novae are detected by {\it Fermi} and which are not.  The novae in the unshaded region to the lower left are almost all detected, and the novae in the heavily shaded region to the upper right are never detected.  The novae in the lightly-shaded middle region are where the {\it Fermi} detection limits vary around in the usual ways.

\subsubsection{Conclusions for $\gamma$-Novae}

The $F_{\gamma}$ correlates with the nova distance and its proxy $V_{\rm peak}$, with the cause being the simple inverse square law.  I have only found one correlation involving intrinsic properties, and that is a moderately loose correlation between $L_{\gamma}$ and $t_3$.  I do not understand the mechanism behind this correlation.  I hope that this is a clue for modelers and theorists.

Previously, our community has the strong expectation that the $\gamma$ emission arises from the shocks from the nova ejecta ramming into other gases, either into circumbinary mass in a stellar wind from the companion (external shocks) or into earlier shells of slower velocity (internal shocks).  This has the strong argument that pion-production from the shocks is the only way that theorists have for making $\gamma$-rays in bulk.  A second strong argument is that V906 Car has its optical and $\gamma$-ray light curve following tightly through three simultaneous optical jitters (flares) and $\gamma$-ray flares (Aydi et al. 2020).  For this second argument, an unproven presumption is that the optical jitters are caused by internal shocks, typical of J-class novae.

I have tested the association between shocks and $\gamma$-emission in a number of ways, and the association fails badly in all cases:  {\bf (A)} All four of the other J-class $\gamma$-novae show either no-correlation or anti-correlation between optical jitters and $\gamma$-ray flares.  So the strong evidence for shock-generation from V906 Car is broadly not applicable for most $\gamma$-novae.  {\bf (B)} If external shocks are significant, then the novae with red giant companions must be greatly more luminous than all other novae, because only the red giants produce massive stellar winds.  Contrarily, the novae with red giant companions are not significantly more luminous than the other systems, so external shocks can at most produce small $\gamma$-ray luminosity.  {\bf (C)} If external shocks are significant, then the systems with the densest stellar winds should produce the brightest $L_{\gamma}$.  V407 Cyg has a Mira star companion, and this will have a stellar wind that is greatly denser than those from the other four $\gamma$-nova with red giant companions.  Contrarily, V407 Cyg has $L_{\gamma}$ fainter than all of the other four systems.  So external shocks are apparently providing little of the $\gamma$-radiation.  {\bf (D)} If  discrete internal shocks dominate, then we have a hard time explaining how most optical light curves can be so smooth.  That is, if internal shocks dominate, then they should manifest in the optical as a superposition of days-long jitters.  Without careful coordination of the timing and strengths of the many individual jitters, a smooth light curve is impossible.  So discrete internal shocks cannot provide any significant light for the optical light curve, and nothing for the $\gamma$-ray light curve.  {\bf (E)}  For internal and external shocks, $L_{\gamma}$ must be proportional to $M_{\rm ejecta}$$FHWM^2$.  Contrarily, there is no correlation or trend between $L_{\gamma}$ and FWHM.  Further, contrarily, $L_{\gamma}$ is not correlated with $M_{\rm WD}$ or the light curve class, as proxies for the ejected mass.

So we have two strong arguments for the $\gamma$-ray emission being associated with shocks from the nova ejecta.  And we have all the experimental evidence (other than for V906 Car) showing that the $\gamma$-ray emission is {\it not} associated with shocks.  I do not know how to resolve this case.

We can make a useable prediction of $\gamma$-nova visibility with {\it Fermi}, as based on the $D$ and $t_3$ measures.  Equation 6 allows us to estimate the nova's $F_{\gamma}$, and for {\it Fermi} thresholds that vary from 0.8--5.0 (in units of $10^{-7}$ photons cm$^{-2}$ s$^{-1}$), we can predict the detectability (see Figure 10).  It turns out that roughly 20\% of discovered novae are detectable by {\it Fermi}.

We can answer the question as to whether the $\gamma$-novae form some sort of distinct class of novae.  Such would be recognized by some intrinsic property or properties forming a bimodal distribution.  The correlation analyses from Section 3.10.1 demonstrate that all the fundamental properties of $\gamma$-novae and non-$\gamma$-novae are indistinguishable.  Figure 10 shows that the $\gamma$-novae are just an intermixed population that forms a continuum in $L_{\gamma}$.  From this, I conclude that {\it all} novae are $\gamma$-ray emitters, with {\it Fermi} only happening to detect the most luminous and nearby examples.

\section{$M_{\rm WD}$ AS THE PRIMARY DETERMINANT OF NOVA PROPERTIES}

\begin{figure*}
	\includegraphics[width=2.1\columnwidth]{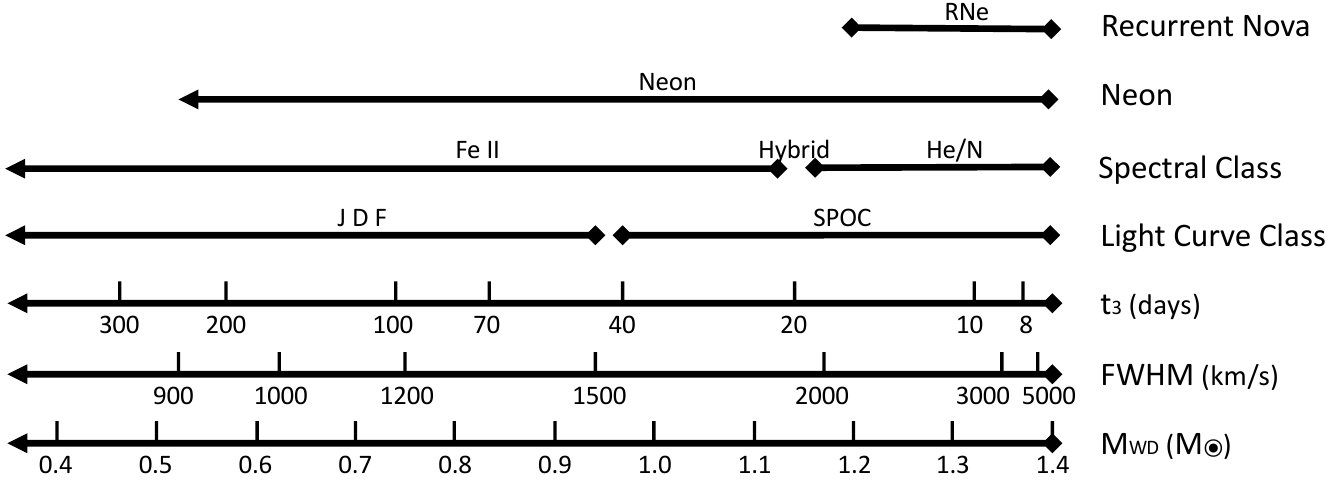}
    \caption{Table 1 contains the primary data product of this paper, with the comprehensive catalog of nova properties (including $M_{\rm WD}$ and $V_q$) for all 402 Galactic novae, while this figure summarizes the science results from this paper.  This figure works with a vertical line intersecting the various number lines to show a possible or likely property for any single nova.  For example, a vertical line that crosses the bottom number line at 0.7 $M_{\odot}$ points to a nova with a FWHM of 1100 km s$^{-1}$, $t_3$ near 120 days, a light curve shape from the J, D, or F classes, an Fe II spectral class, possibly a neon nova, and certainly not a recurrent nova.  A known RN will have a mass 1.2--1.4 $M_{\odot}$, FWHM 2000--6000 km $^{-1}$, $t_3$ 6--18 days, a light curve shape from the S, P, O, or C classes, a He/N spectral class, and possibly be a neon nova.  The $t_3$ number line is related to the WD mass by Equation 2, the FWHM is from Equation 4, the division between JDF and SPOC classes is near 0.95 $M_{\odot}$, and the division between Fe II and He/N spectral classes is near 1.15 $M_{\odot}$.  These divisions and equations are consistent with being exact relations with substantial superposed measurement errors.  Nevertheless, there could easily be real scatter and outliers to some modest degree.}
\end{figure*}

The primary science result from this paper is to quantify the dependencies on $M_{\rm WD}$ for many of the fundamental observed properties on novae. These can be represented by number lines, as in Figure 11.  I find that $t_3$ equaling $10^{-1.73M_{\rm WD}}$$\times$1900 days, with high-mass meaning fast light curves.  The FWHM is well-represented by FWHM = 0.23$\times V_{\rm escape}$, although there is an upturn for $>$1.3 $M_{\odot}$.  The WD masses for light curve classes S, P, O, and C are indistinguishable, the masses for J, D, and F classes are indistinguishable, but there appear to be a sharp division with all (or almost all) SPOC novae $>$0.97 $M_{\odot}$ and JDF novae with $<$0.97 $M_{\odot}$.  The WD mass is also the primary determinant of the spectral class, with Fe II novae for $<$1.15 $M_{\odot}$, He/N novae for $>$1.15 $M_{\odot}$, and hybrid novae around the division.  Thus, $M_{\rm WD}$ has the primary control over light curves and spectra of nova eruptions.

The relations and divisions from the previous paragraph all display substantial scatter (see Figures 3, 4, and 5, plus Table 3).  All, or almost all, of this scatter is expected from the known measurement errors.  Section 2.1 goes to length to expose the large real measurement uncertainties, where the one-sigma error bar in $M_{\rm WD}$ is $\pm$0.15 $M_{\odot}$.  With this, we must see the SPOC mass distribution with a half-Gaussian tail below 0.97 $M_{\odot}$, and the JDF having a tail above 0.97 $M_{\odot}$.  And the scatter in Figure 4, around the best fit line of Equation 2, is much as expected from measurement error in the WD masses.  For all four of the equations and divisions, the observations are consistent with the equations and divisions being exact and universal.

The equations and divisions might be exact, or some additional effects might produce some other source of scatter lost within the measurement errors.  For example, some special case or condition might lead to outliers that violate the equations and divisions.  The unique helium nova V445 Pup provides a perfect example, where the SPOC-division and the $t_3$ equation are both greatly violated, simply because the physics of this helium-burning nova is greatly different from all the other hydrogen-burning novae.  It is easy to imagine other possibilities to create outliers.  And there are likely non-dominant effects that will make for some real scatter about the equations and divisions.  For example, I expect that there is some dependency of FWHM on the orbital period, and such could induce a real horizontal scatter in Figure 5 and inexactitude with Equation 4.  Still, given that the observed scatter is close to that expected from ordinary measurement errors, the underlying equations and divisions can only have additional sources of scatter that are relatively small.  That is, the equations and divisions must be pretty good, with at most small intrinsic errors.

The WD mass does {\it not} have useful correlations with neon novae, the disk versus bulge populations, the orbital period, nova shells, $\gamma$-ray emission, or the peak absolute magnitude.  That neon novae WDs have large fractions with $<$1.2 or $<$1.0 $M_{\odot}$ provides a demonstration that these nova are all losing mass over each eruption cycle, plus a likely conclusion that all novae (and all CVs) are having their WDs being whittled down in mass over long time scales.  I am surprised by the lack of correlation with shells, as I expect that low-mass WDs would create brighter shells.  I am surprised by the lack of correlation with $\gamma$-ray emission, as I expect that high-mass WDs would have higher shock velocities.  There are correlations with orbital period ($P$$>$100 day systems are all near the Chandrasekhar mass, and the $P$$<$0.7 day systems have nearly no high-mass WDs), but these are complicated for evolution considerations due to discovery selection effects.

A variety of follow-on work is needed.  I would like to construct and collect comprehensive lists of pre-eruption magnitudes, accretion rates, and post-eruption long-term decline rates for all possible novae.  My new lists of nova periods and masses should be combined with similar lists for CVs of all types so as to test whether the WD mass is increasing or decreasing over time.   Theorists should construct models that reproduce the observed behavior for $t_3$ (see Figure 4 and Equation 2) and the FWHM (see Figure 5 and Equation 4).  The FWHM work must include calculations of realistic line profiles for comparison with the model velocity distributions.  It would be nice to get an explanation for why the average FWHM increases much faster than the escape velocity as the Chandrasekhar mass is approached (see Figure 5).  Theorists need to come up with plausible and testable ideas to explain the light curve cusps, oscillations, jitters, and flat-tops.  Theorists should explain the visibility and brightness of nova shells and late-time $\gamma$-ray emission.

\begin{acknowledgments}
I thank M. M. Shara (American Museum of Natural History) for discussions on the science.
\end{acknowledgments}

%



{}


\end{document}